\documentclass[submission, Phys]{SciPost}

\usepackage{amsmath, amssymb}
\usepackage{graphicx}
\usepackage{hyperref}
\usepackage{booktabs}
\usepackage{xcolor}
\usepackage[normalem]{ulem} 
\usepackage{braket}
\usepackage{tikz}
\usepackage{soul}
\usepackage{subcaption}
\usetikzlibrary{positioning, arrows.meta, fit,
  decorations.pathreplacing, calc, patterns}
\usepackage{comment}
\usepackage[normalem]{ulem}
\usepackage{xcolor}
\usepackage{soul}   
\usepackage[table]{xcolor}
\definecolor{tablegray}{gray}{0.94}
\definecolor{boxgray}{gray}{0.97}
\sethlcolor{yellow}

\hypersetup{
  colorlinks,
  linkcolor={red!50!black},
  citecolor={blue!50!black},
  urlcolor={blue!80!black}
}

\begin{document}

\begin{center}
{\Large \textbf{\boldmath
Vectorized Symmetric and Fermionic Tensor Network Implementations
for GPU-Accelerated Variational Monte Carlo
}}

Si-Jing Du$^{1,*}$,
Johnnie Gray$^{2,*}$,
Ao Chen$^{2}$,
Garnet Kin-Lic Chan$^{2,\dagger}$

\vspace{6pt}
$^{1}$ Division of Engineering and Applied Science,
California Institute of Technology, Pasadena, CA 91125, USA

$^{2}$ Division of Chemistry and Chemical Engineering,
California Institute of Technology, Pasadena, CA 91125, USA

\vspace{4pt}
{\small $*$ These authors contributed equally} \\
{\small $\dagger$ Corresponding author: gkc1000@gmail.com}

\vspace{4pt}
\textbf{Abstract}
\end{center}

\noindent

Variational Monte Carlo (VMC) calculations based on tensor networks (TN) have
recently achieved competitive accuracy in ground-state calculations of strongly correlated spin and fermionic systems. However, existing tensor network VMC (TN-VMC) algorithms have not been formulated in a manner that can fully utilize GPU acceleration.
We tackle the key missing ingredient, namely, the vectorized evaluation of
tensor network amplitudes and tensor network operations. This ensures high GPU utilization by batching over computations with identical structure. In particular, we show how to achieve vectorization for the practically relevant case of abelian symmetric tensor networks (which includes fermionic tensor networks) by
developing a ``flat'' tensor network formalism for block-sparse tensor
representation and contraction. Using this, we construct a GPU-adapted symmetric TN-VMC workflow with batched 
tensor network computation.
In the two-dimensional Fermi--Hubbard model, we demonstrate a GPU speedup of
up to $300 \times$ over single core CPU implementations, for fermionic TN variational wavefunctions. 


\vspace{10pt}
\noindent\rule{\textwidth}{0.4pt}
\tableofcontents
\noindent\rule{\textwidth}{0.4pt}
\vspace{10pt}

\section{Introduction}
\label{sec:intro}

Variational Monte Carlo (VMC) is a powerful approach for ground-state
calculations of quantum many-body
systems~\cite{mcmillan1965ground,ceperley1977monte,foulkes2001quantum}.
Among the variational ans\"atze proposed for VMC, tensor network
states (TNS)~\cite{Pi_orn_2010,Kraus_2010,Corboz_2010,
gu2010grassmanntensornetworkstates,Singh_2010,Singh_2011,Bauer_2011}
have achieved competitive accuracy in ground-state calculations of
two-dimensional strongly correlated
systems~\cite{Liu_2017,Liu2021accuratepeps, fan2026nntns}, including for fermionic problems~\cite{liu2025fpeps,du2024neuralized,Robledo_Moreno_2022}.

\sloppy
Despite this progress, large-scale tensor network VMC (TN-VMC)
calculations have remained predominantly
CPU-based~\cite{Liu_2017,
liu2024tensor,liu2025fpeps,du2024neuralized}.
This contrasts with other large-scale tensor network workloads that already
benefit substantially from GPU acceleration, including quantum-circuit
simulation~\cite{Begu_i__2024,pan2024efficientquantumcircuitsimulation,
pancircuit2022,Pan_2022,Huang_2021nsy,Pan_2020,Gray2021hyperoptimized,
Gray2024hyperoptimized} and tensor network calculations of partition
functions~\cite{Liu_2021_tropicaltensor,Liu_2023_combinatorialoptimization,
chen2025batchtnmcefficientsamplingtwodimensional,
Gray2021hyperoptimized,Gray2024hyperoptimized}.
Enabling TN-VMC to efficiently utilize GPUs, however, requires more than simply offloading the tensor contractions to GPUs. A VMC calculation repeatedly evaluates wavefunction amplitudes for a large number of physical configurations, and these evaluations are mutually independent at each sampling step. To exploit the high-throughput parallelism of modern GPUs, these amplitudes must therefore be evaluated concurrently.
Such parallelization is typically realized through vectorization of the wavefunction evaluation. In standard TN-VMC, evaluating the wavefunction for a given configuration involves constructing and contracting a single-layer amplitude tensor network. An efficient GPU implementation must therefore construct and contract such amplitude networks, vectorized over batches of configurations.

In this work, we describe how this can be achieved. For TNS without symmetry, where all tensors are dense arrays, vectorization is straightforward. However, many TN calculations either benefit from symmetry (to reduce storage and speedup computation), or intrinsically require symmetry in their formulation. A key example of the latter is the fermionic TNS (fTNS) formalism, where every tensor is chosen to be $\mathbb{Z}_2$ symmetric to locally implement fermionic anticommutation rules.~\cite{gu2010grassmanntensornetworkstates,Corboz_2010,Pi_orn_2010,Kraus_2010,gao2024fermionic, Mortier_2025} The block-sparse tensor network structure associated with the symmetries introduces obstacles to efficient vectorization.

To tackle this problem, we describe 
a \textit{flat symmetric tensor computational formalism} that supports vectorization for finite cyclic symmetries (e.g. the $\mathbb{Z}_2$ symmetry associated with fermions), and thus, an efficient GPU-accelerated TN-VMC workflow. Our idea is related to previous work by some of us on the \texttt{Symtensor} library~\cite{SciPostPhysCodeb.10}, but generalizes that work (which was limited to tensor contractions) to the full set of operations required for VMC, and in particular, the additional complications arising from tensor slicing for amplitude evaluation.

To demonstrate the performance of our approach, particularly for symmetric tensors, we use
 the two-dimensional Fermi--Hubbard model, where we assess
GPU speedups for two classes of fTNS: standard fermionic
projected entangled-pair states (fPEPS) and a tensor network function ansatz
based on approximate fPEPS contraction (TNF-fPEPS). Our results
demonstrate that the GPU implementation achieves {considerable} 
speedups over conventional MPI-parallelized CPU implementations in
production-scale VMC calculations.
Beyond these performance gains, the GPU-accelerated TN-VMC workflow also enables the efficient optimization of the recently introduced neuralized fermionic tensor network state (NN-fTNS)~\cite{du2024neuralized}, which incorporates a deep neural-network component. Our preliminary results show that a deep NN-fTNS can achieve ground-state energies competitive with
recently reported results from large-scale fermionic neural quantum states (NQS)~\cite{gu2025solvinghubbardmodelneural} and
conventional fPEPS with substantially larger bond dimensions~\cite{liu2025fpeps}.

The calculations in this work utilized several libraries. General tensor network construction, algorithms, and operations are orchestrated by \texttt{quimb}~\cite{gray2018quimb}; the flat symmetric tensor formalism 
is implemented in a new library \texttt{symmray}~\cite{gao2024fermionic,symmray}; and the VMC calculations were carried out using \texttt{vmc\_torch}~\cite{Du_vmc_torch_Flexible_Variational}, a PyTorch-based
library~\cite{torch2019github}, and \texttt{Quantax}~\cite{quantax}, a GPU-optimized VMC
framework based on JAX~\cite{jax2018github}.


The remainder of this paper is organized as follows. In
Section~\ref{sec:vmc}, we briefly review the variational Monte Carlo method and
explain the role of vectorized amplitude evaluation in
GPU-accelerated VMC. Section~\ref{sec:vectorization} presents the vectorized
tensor network amplitude formulation. We first discuss the dense
tensor network case and then introduce the flat symmetric tensor formalism,
which enables vectorized amplitude evaluation for abelian-symmetric TN, such as fTNS. In
Section~\ref{sec:results}, we present numerical benchmarks on the
two-dimensional Fermi--Hubbard model. Section~\ref{sec:vmc_torch} briefly describes
the software stack. Finally, we summarize
our work and future directions in Section~\ref{sec:summary}.

\section{Variational Monte Carlo}
\label{sec:vmc}

\subsection{VMC formulation}
\label{sec:vmc_formulation}

We briefly review the VMC framework and establish the notation used in the rest
of the paper. In a typical ground-state VMC calculation, a parameterized trial
wavefunction $\ket{\Psi_\theta}$ is optimized to minimize the variational
energy and approximate the ground state of a many-body Hamiltonian $H$.
Consider $H$ represented in a product-state basis $\{\ket{\sigma}\}$, where
$\sigma=(\sigma_1,\ldots,\sigma_N)$ labels a physical configuration. The
variational energy is calculated as
\begin{equation}
  E(\theta)
  =
  \frac{\braket{\Psi_\theta|H|\Psi_\theta}}
       {\braket{\Psi_\theta|\Psi_\theta}}
  =
  \mathbb{E}_{\sigma\sim p_\theta}
  \left[
    E_{\mathrm{loc}}(\sigma)
  \right],
  \label{eq:vmc_energy}
\end{equation}
where configurations $\sigma$ are sampled from the Born distribution
\begin{equation}
  p_\theta(\sigma)
  =
  \frac{|\Psi_\theta(\sigma)|^2}
       {\sum_{\sigma'}|\Psi_\theta(\sigma')|^2},
\end{equation}
and the local energy is
\begin{equation}
  E_{\mathrm{loc}}(\sigma)
  =
  \sum_{\sigma'}
  H_{\sigma\sigma'}
  \frac{\Psi_\theta(\sigma')}{\Psi_\theta(\sigma)}.
  \label{eq:local_energy}
\end{equation}
Here $\Psi_\theta(\sigma)=\braket{\sigma|\Psi_\theta}$ is the wavefunction
amplitude in the chosen basis.

In practice, configurations are commonly generated by sampling from $p_\theta$ via
Markov-chain Monte Carlo based on the
Metropolis--Hastings algorithm~\cite{Metropolis_1953,Hastings_1970}. 
For symmetric proposal moves, a proposed
configuration $\sigma'$ is accepted with probability
\begin{equation}
  A(\sigma\to\sigma')
  =
  \min\left[
    1,
    \left|
      \frac{\Psi_\theta(\sigma')}
           {\Psi_\theta(\sigma)}
    \right|^2
  \right].
  \label{eq:metropolis_acceptance}
\end{equation}

To minimize the variational energy $E(\theta)$, the wavefunction parameters $\theta$ are
iteratively updated using gradient descent methods. At each VMC step, we compute the derivatives of the sampled wavefunction amplitudes with respect to $\theta$. From these
derivatives, we obtain the parameter update $\delta\theta$ direction via
optimization algorithms based on stochastic reconfiguration
(SR)~\cite{sorella1998green,chen2024minsr}, and the parameters are updated as
$\theta \leftarrow \theta+\eta\cdot\delta\theta$ where $\eta>0$ is the update step size.

For sampling, the acceptance probability in Eq.~\eqref{eq:metropolis_acceptance} has the form of a ratio between the amplitude of the proposed configuration and current configuration. For local-energy
evaluation, Eq.~\eqref{eq:local_energy} requires the amplitudes
$\Psi_\theta(\sigma')$ of configurations connected to each sampled
configuration $\sigma$ by the Hamiltonian. Therefore, the main parts of the VMC loop
involve repeated calls to the amplitude function $\Psi_\theta(\sigma)$.

\begin{figure}[t]
\centering
\includegraphics[width=1.05\linewidth]{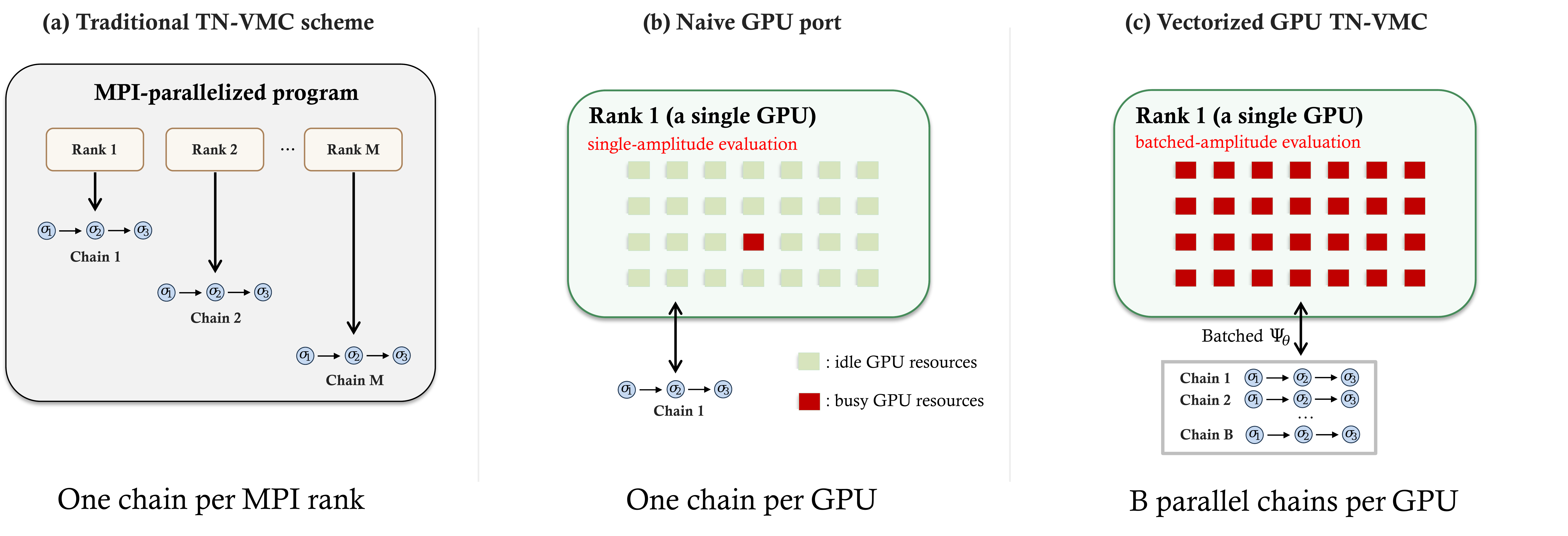}
\caption{\textbf{Parallelization schemes for TN-VMC sampling.}
(a) In traditional TN-VMC, an MPI-parallelized program assigns one Markov
chain to each MPI rank, and amplitudes are evaluated sequentially along
each chain. (b) A naive GPU port keeps the same one-chain-per-rank
structure, exposing only a single-configuration amplitude
evaluation to the GPU at each Markov move. (c) The vectorized GPU scheme
assigns $B$ chains to each GPU rank and evaluates the corresponding amplitudes
at each move stage as a batched-amplitude evaluation with a shared
computational graph and a leading batch axis.}
\label{fig:vmc_parallel_schemes}
\end{figure}

\subsection{Parallelism in VMC}
\label{sec:vmc_parallelism}

The large number of wavefunction-amplitude evaluations typically constitutes the dominant computational cost in VMC. This cost can be reduced through parallel and distributed computing.

VMC naturally inherits the embarrassingly parallel structure of Monte Carlo sampling. Independent Markov chains can be distributed across MPI processes for parallel configuration sampling, while local observables measured on different chains can be combined through MPI collective operations. Within each individual Markov chain, however, the sampling procedure is inherently sequential: the chain evolves through a sequence of local-move proposals, and each accept--reject decision depends on the current configuration of that chain.

In conventional CPU-based TN-VMC implementations on computing clusters [Fig.~\ref{fig:vmc_parallel_schemes}(a)], one typically assigns a single Markov chain to each MPI rank, with each rank mapped to a hardware compute unit consisting of, for example, a small number of CPU cores. This coarse-grained MPI parallelism has been the standard parallelization strategy adopted in previous TN-VMC studies~\cite{Liu_2017,liu2025fpeps}.

However, the one-chain-per-rank strategy described above becomes inefficient when VMC is executed on GPUs. On GPU clusters, an MPI-parallel program typically assigns one entire GPU device to each MPI rank. Within a single Markov chain, each proposed move requires only one amplitude evaluation for the proposed configuration in order to perform the accept--reject test. If only one chain is assigned to each GPU, amplitudes on that GPU are evaluated for individual configurations one at a time as the chain evolves sequentially. For the tensor network ans\"atze considered in this work, the computational work for the contraction associated with a single-configuration amplitude evaluation is typically too small to saturate a modern GPU [Fig.~\ref{fig:vmc_parallel_schemes}(b)]. Consequently, a direct GPU port of a conventional TN-VMC implementation does not by itself ensure efficient GPU utilization.

A GPU-efficient VMC implementation must instead assign multiple Markov chains to each GPU [Fig.~\ref{fig:vmc_parallel_schemes}(c)], as is standard in modern GPU-oriented VMC frameworks with other kinds of wavefunctions~\cite{Filippo2022netket,Schmitt_2022}. For a batch size $B$, a single GPU advances $B$ independent chains concurrently. Although the Markov updates within each individual chain remain sequential, the same update stage is performed across all $B$ chains in parallel. Specifically, the proposed configurations from the $B$ chains are assembled into a batch, and their amplitudes are evaluated in a single vectorized operation. This fine-grained parallelization increases the computational workload of each GPU operation and thereby enables more effective use of the device's parallel throughput.

More generally, whenever amplitudes must be evaluated for a large number of configurations, a GPU can process them efficiently in batches of size $B$. In the ideal limit of perfect parallel scaling, this batched evaluation can provide a speedup of up to a factor of $B$ relative to sequential evaluation, although the actual speedup is generally limited by hardware utilization, memory traffic, and vectorization overhead.

\subsection{Vectorized amplitude evaluation in GPU-adapted VMC}
\label{sec:vectorized_vmc}

The discussion above shows that GPU-adapted VMC requires a wavefunction ansatz that supports batched amplitude evaluation. For an individual physical configuration, the wavefunction defines the scalar map
\begin{equation}
\Psi_\theta:
\sigma \longmapsto \Psi_\theta(\sigma).
\end{equation}
Vectorized wavefunction evaluation replaces this scalar map with the batched map
\begin{equation}
\vec{\Psi}\theta:
(\sigma_1,\ldots,\sigma_B)
\longmapsto
\left(
\Psi_\theta(\sigma_1),
\ldots,
\Psi_\theta(\sigma_B)
\right),
\label{eq:vectorized_wavefunction}
\end{equation}
where $B$ is the leading batch dimension that denotes the number of configurations processed in a single vectorized amplitude-evaluation call.

For efficient GPU execution, the batched map in Eq.~\eqref{eq:vectorized_wavefunction} should be implemented as a dense batched tensor computation. This requires all $B$ amplitude evaluations to execute the \textbf{same computational graph}, such that after vectorization, each intermediate tensor only needs to carry an extra leading batch dimension $B$ and each operation is replaced by its batched counterpart. If the computational graph or intermediate tensor shapes depend on the physical configuration, the batch must instead be partitioned into smaller groups or processed sequentially, thereby reducing GPU efficiency.

\section{Vectorized tensor-network amplitude evaluation}
\label{sec:vectorization}

\subsection{Tensor networks and amplitudes}
\label{sec:tn_amplitudes}

A tensor network state represents a many-body wavefunction by assigning local
tensors to the sites of a lattice and contracting them along shared auxiliary
bonds~\cite{Ran_2020,Or_s_2014}. In a product basis
$\ket{\sigma}=\ket{\sigma_1,\ldots,\sigma_N}$, the state is written as
\begin{equation}
  \ket{\Psi}
  =
  \sum_{\sigma}
  \Psi(\sigma)\ket{\sigma}.
\end{equation}
We will assume that each site is associated with a single tensor (although the analysis can be generalized to cases where there are extra tensors not associated with sites). Consequently, every tensor has a physical index associated with its site, and the tensor network defining the amplitude can be viewed as a `single-layer'.

For a given physical configuration
$\sigma=(\sigma_1,\ldots,\sigma_N)$, the amplitude
$\Psi(\sigma)=\braket{\sigma|\Psi}$ is obtained by fixing the physical index of
each site tensor to the corresponding local value $\sigma_i$ and contracting
the resulting tensor network. Schematically, the amplitude can be
written as
\begin{equation}
  \Psi(\sigma)
  =
  \sum_{\{\alpha_e\}_{e\in E}}
  \prod_{i=1}^{N}
  T^{[i],\,\sigma_i}_{\{\alpha_e:\,e\ni i\}},
  \label{eq:tn_amplitude}
\end{equation}
where $E$ denotes the set of auxiliary bonds in the tensor network and
$\alpha_e$ denotes the auxiliary index associated with bond $e$. The notation
$\{\alpha_e:\,e\ni i\}$ denotes the collection of auxiliary indices attached
to site $i$, while the sum over $\{\alpha_e\}_{e\in E}$ contracts all shared
auxiliary indices. The auxiliary bond dimensions control both the number of
variational parameters and the expressive power of the ansatz. Matrix product
states (MPS) in one dimension and projected entangled-pair states (PEPS) in two
dimensions are standard examples of tensor network states~\cite{white1992density,
Or_s_2014,schollwock2011density,verstraete2004renormalization,Cirac_2021}.

The discussion below applies to any definition of the tensor-network amplitude map
\begin{equation}
  \sigma \longmapsto \Psi(\sigma)
\end{equation}
that includes tensor network contraction. 
Besides conventional tensor network states, this also includes more general tensor network functions (TNFs)~\cite{liu2024tensor} that allow 
for non-linear operations in the definition of the map. 
We consider a representative fermionic TNF, the NN-fTNS~\cite{du2024neuralized}, later in this work. 

\subsection{Requirements for vectorized tensor-network amplitude evaluation}
\label{sec:tn_vectorizability}

Consider a batch of physical configurations
$\{\sigma^{(1)},\ldots,\sigma^{(B)}\}$. For each configuration
$\sigma^{(a)}$, the tensor network amplitude is evaluated by first selecting,
at every site $i$, the tensor slice corresponding to the local physical index
$\sigma_i^{(a)}$ and then contracting the resulting single-layer amplitude
network to a scalar. The batched amplitude-evaluation problem therefore
consists of the $B$ maps
\begin{equation}
  \sigma^{(a)}
  \longmapsto
  \Psi\!\left(\sigma^{(a)}\right),
  \qquad
  a=1,\ldots,B.
\end{equation}

These $B$ amplitude evaluations can be combined into a single vectorized
tensor-network computation only if they admit a common computational graph.
Here, a common computational graph for tensor network amplitudes means that, after fixing the physical
indices, all single-layer tensor networks in the batch are contracted using the same
sequence of tensor operations, with compatible intermediate tensor shapes and
an identical contraction path~\cite{Gray2021hyperoptimized}. 

The common-graph condition is automatically satisfied for conventional tensor
networks composed of dense tensors, as we briefly explain in
Sec.~\ref{sec:dense_tn}. For symmetric tensor networks (such as fermionic tensor networks) represented by
block-sparse tensors, however, the situation is more subtle, and in general, symmetric TNS do not share a common amplitude computational graph for different input configurations, which we explain in Sec.~\ref{sec:fermionic}.

\subsection{Dense tensor-network amplitudes}
\label{sec:dense_tn}

For tensor networks composed of dense tensors, vectorized amplitude evaluation
follows directly from the dense tensor representation. Consider a local tensor
$T^{[i]}$ with one physical index of dimension $d$ and $z_i$ auxiliary bond
indices with dimensions $D_1,\ldots,D_{z_i}$,
\begin{equation}
  T^{[i]} \in \mathbb{C}^{D_1 \times \cdots \times D_{z_i} \times d},
\end{equation}
where $z_i$ is the number of auxiliary bonds incident to site $i$. For a single
configuration $\sigma$, projecting the physical leg at site $i$ amounts to
selecting the slice
\begin{equation}
  A^{[i]}(\sigma_i)
  =
  T^{[i]}_{:,\ldots,:,\sigma_i}
  \in
  \mathbb{C}^{D_1 \times \cdots \times D_{z_i}} .
  \label{eq:bosonic_single_slice}
\end{equation}
The amplitude $\Psi(\sigma)$ is then obtained by contracting the single-layer
network formed from the projected tensors $\{A^{[i]}(\sigma_i)\}_{i=1}^{N}$.

For a batch of configurations $\{\sigma^{(1)},\ldots,\sigma^{(B)}\}$, the
corresponding projected site tensors can be collected into a batched tensor
\begin{equation}
  A^{[i]}_B
  =
  \operatorname{stack}_{a=1}^{B}
    A^{[i]}(\sigma_i^{(a)})
  \in
  \mathbb{C}^{B \times D_1 \times \cdots \times D_{z_i}}.
  \label{eq:bosonic_batched_slice}
\end{equation}
Thus, batching introduces only a leading batch dimension, while the auxiliary
tensor shapes and the contraction pattern remain identical across all
configurations.

The original single-configuration contraction can therefore be lifted directly
to a batched contraction. Each projected site tensor carries the same leading
batch dimension, which is preserved throughout all tensor contractions. The
final contraction yields the vector of amplitudes
\begin{equation}
  \vec{\Psi}
  =
  \left(
    \Psi(\sigma^{(1)}),
    \ldots,
    \Psi(\sigma^{(B)})
  \right).
\end{equation}
Hence, amplitude evaluation for dense tensor networks is naturally
vectorizable.

\subsection{Symmetric tensor network amplitudes}
\label{sec:fermionic}

\subsubsection{Symmetry-resolved block-sparse tensors and configuration
dependence in amplitude evaluation}

For states that transform as irreps (`fixed charge') under a global symmetry, it is standard to choose a symmetric tensor network ansatz where the individual tensors are charge conserving. This induces a block-sparse structure in the tensors following charge conservation, which leads to new considerations.


More specifically, we will restrict ourselves to abelian symmetry and
adopt the notation of
Ref.~\cite{SciPostPhysCodeb.10} to describe block-sparse tensors with such symmetry. Each tensor index is decomposed into a charge-sector label
and an intra-sector index. We write such an index as $iI$, where the uppercase
label $I$ denotes the charge sector and the lowercase label $i$ indexes the
degeneracy space within that sector. A rank-$k$ tensor is therefore written as
$T_{i_1 I_1,\ldots,i_k I_k}$.

For an abelian symmetry group $G$, a tensor element is nonzero only when its
charge labels satisfy a conservation rule. Assign to each leg $\ell$ an
orientation $\epsilon_\ell\in\{+1,-1\}$, with a leg and its dual carrying
opposite orientations. We then write
\begin{equation}
  T_{i_1 I_1,\ldots,i_k I_k}
  =
  0
  \qquad
  \text{unless}
  \qquad
  \epsilon_1 I_1+\cdots+\epsilon_k I_k = Q
  \quad \text{in } G,
  \label{eq:fermion_charge_rule}
\end{equation}
where $Q$ is the total charge of the tensor and the addition denotes the group
operation, with $\epsilon_\ell I_\ell$ denoting either $I_\ell$ or its group
inverse according to the leg orientation. Equivalently, the tensor decomposes
into charge-sector blocks,
\begin{equation}
  T
  =
  \bigoplus_{\substack{I_1,\ldots,I_k\\
  \epsilon_1 I_1+\cdots+\epsilon_k I_k = Q \;\mathrm{in}\; G}}
  T^{(I_1,\ldots,I_k)},
  \label{eq:block_decomp}
\end{equation}
where each block $T^{(I_1,\ldots,I_k)}$ is a dense tensor over the corresponding
intra-sector indices,
\begin{equation}
  T^{(I_1,\ldots,I_k)}_{i_1,\ldots,i_k}
  \equiv
  T_{i_1 I_1,\ldots,i_k I_k}.
  \label{eq:block_tensor_definition}
\end{equation}

This block-sparse structure can render symmetric tensor-network amplitude
evaluation configuration dependent. In tensor-network amplitude evaluation,
the physical index of each site tensor is fixed according to the corresponding
local configuration $\sigma_i$. Since a symmetry-resolved tensor index is
written as $iI$, fixing the physical index selects both a physical charge sector
$I_m$ and an intra-sector index $i_m$ on the physical leg $m$. The projected
tensor is therefore obtained in two steps: one first retains only the
charge-sector blocks satisfying
\begin{equation}
  \epsilon_1 I_1+\cdots+\epsilon_{m-1}I_{m-1}
  +\epsilon_m I_m+\epsilon_{m+1}I_{m+1}+\cdots+\epsilon_k I_k
  =
  Q
  \quad \text{in } G,
  \label{eq:projected_charge_rule}
\end{equation}
and then slices the corresponding dense blocks along the $m$-th intra-sector
axis at $i_m$.

The crucial point is that different local physical states may carry different
physical charges $I_m$. Changing $I_m$ changes the set of allowed charge tuples
on the remaining auxiliary legs through
Eq.~\eqref{eq:projected_charge_rule}. Moreover, if the intra-sector dimension
$D^{(\ell)}_{I_\ell}$ of an auxiliary leg $\ell$ depends on its charge label
$I_\ell$, then the dense shape of an allowed projected block,
\begin{equation}
  D^{(1)}_{I_1}
  \times\cdots\times
  D^{(m-1)}_{I_{m-1}}
  \times
  D^{(m+1)}_{I_{m+1}}
  \times\cdots\times
  D^{(k)}_{I_k},
  \label{eq:projected_block_shape}
\end{equation}
also depends on the selected physical charge $I_m$. Consequently, two
configurations with different local physical states can produce projected
tensors containing different sets of charge-sector blocks and different dense
block shapes. The resulting single-layer amplitude networks therefore need not
share compatible tensor shapes or a common contraction computational graph, as required in
Sec.~\ref{sec:tn_vectorizability}. This configuration dependence obstructs the
direct vectorization of amplitude evaluation.

\subsubsection{Flat symmetric tensor formalism}

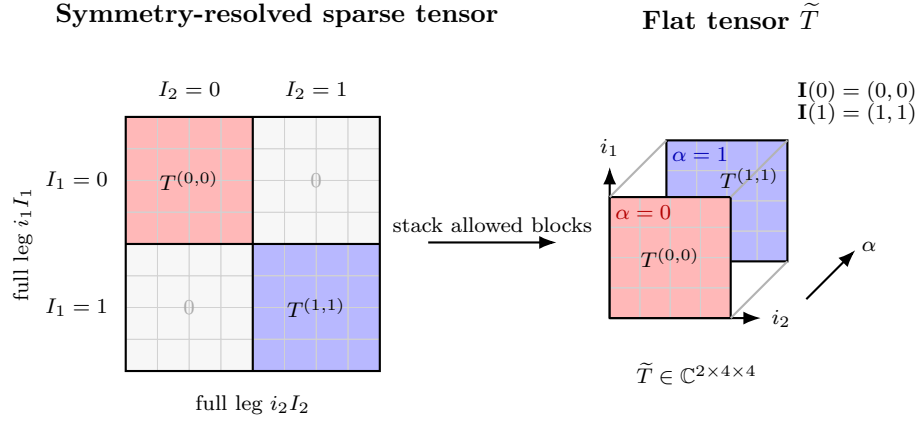
\begin{figure}[!t]
\centering
\begin{tikzpicture}[
  cell/.style={draw=gray!35, line width=0.25pt},
  arrow/.style={->, thick, >=Latex},
  ttl/.style={font=\small\bfseries, align=center},
  ax/.style={font=\scriptsize},
  lab/.style={font=\scriptsize, align=center},
]

\node[ttl] at (2.0, 4.7) {Symmetry-resolved sparse tensor};
\def\x0{0}
\def\y0{0}
\def\s{0.42}

\foreach \r in {0,...,7} {
  \foreach \c in {0,...,7} {
    \draw[cell, fill=gray!6]
      ({\x0+\c*\s},{\y0+\r*\s}) rectangle
      ({\x0+(\c+1)*\s},{\y0+(\r+1)*\s});
  }
}
\fill[red!28] ({\x0+0*\s},{\y0+4*\s}) rectangle ({\x0+4*\s},{\y0+8*\s});
\foreach \r in {4,...,7} {
  \foreach \c in {0,...,3} {
    \draw[cell] ({\x0+\c*\s},{\y0+\r*\s}) rectangle ({\x0+(\c+1)*\s},{\y0+(\r+1)*\s});
  }
}
\fill[blue!28] ({\x0+4*\s},{\y0+0*\s}) rectangle ({\x0+8*\s},{\y0+4*\s});
\foreach \r in {0,...,3} {
  \foreach \c in {4,...,7} {
    \draw[cell] ({\x0+\c*\s},{\y0+\r*\s}) rectangle ({\x0+(\c+1)*\s},{\y0+(\r+1)*\s});
  }
}
\draw[thick] ({\x0},{\y0}) rectangle ({\x0+8*\s},{\y0+8*\s});
\draw[thick] ({\x0+4*\s},{\y0}) -- ({\x0+4*\s},{\y0+8*\s});
\draw[thick] ({\x0},{\y0+4*\s}) -- ({\x0+8*\s},{\y0+4*\s});
\node[lab] at ({\x0+2*\s},{\y0+6*\s}) {$T^{(0,0)}$};
\node[lab] at ({\x0+6*\s},{\y0+2*\s}) {$T^{(1,1)}$};
\node[lab, text=gray!65] at ({\x0+6*\s},{\y0+6*\s}) {$0$};
\node[lab, text=gray!65] at ({\x0+2*\s},{\y0+2*\s}) {$0$};
\node[ax, anchor=south] at ({\x0+2*\s},{\y0+8*\s+0.10}) {$I_2=0$};
\node[ax, anchor=south] at ({\x0+6*\s},{\y0+8*\s+0.10}) {$I_2=1$};
\node[ax, anchor=east] at ({\x0-0.10},{\y0+6*\s}) {$I_1=0$};
\node[ax, anchor=east] at ({\x0-0.10},{\y0+2*\s}) {$I_1=1$};
\node[ax, anchor=north] at ({\x0+4*\s},{\y0-0.18}) {full leg $i_2 I_2$};
\node[ax, rotate=90, anchor=south] at ({\x0-1.1},{\y0+4*\s}) {full leg $i_1 I_1$};

\draw[arrow] (4.0,1.7) -- node[above, font=\scriptsize] {stack allowed blocks} (5.7,1.7);

\node[ttl] at (8.0, 4.7) {Flat tensor $\widetilde{T}$};
\coordinate (O) at (6.4,0.7);
\def\uX{0.40}
\def\uY{0.40}
\draw[arrow] (O) -- ++(2.0,0) node[ax, anchor=west] {$i_2$};
\draw[arrow] (O) -- ++(0,2.0) node[ax, anchor=south] {$i_1$};
\begin{scope}[shift={(0.75,0.75)}]
  \fill[blue!28] (6.4,0.7) rectangle ++(1.6,1.6);
  \foreach \r in {0,...,3} {
    \foreach \c in {0,...,3} {
      \draw[cell] ({6.4+\c*\uX},{0.7+\r*\uY}) rectangle ({6.4+(\c+1)*\uX},{0.7+(\r+1)*\uY});
    }
  }
  \draw[thick] (6.4,0.7) rectangle ++(1.6,1.6);
\end{scope}
\fill[red!28] (6.4,0.7) rectangle ++(1.6,1.6);
\foreach \r in {0,...,3} {
  \foreach \c in {0,...,3} {
    \draw[cell] ({6.4+\c*\uX},{0.7+\r*\uY}) rectangle ({6.4+(\c+1)*\uX},{0.7+(\r+1)*\uY});
  }
}
\draw[thick] (6.4,0.7) rectangle ++(1.6,1.6);
\draw[thick, gray!60] (8.0,0.7) -- (8.75,1.45);
\draw[thick, gray!60] (8.0,2.3) -- (8.75,3.05);
\draw[thick, gray!60] (6.4,2.3) -- (7.15,3.05);
\node[lab] at (7.2,1.5) {$T^{(0,0)}$};
\node[lab] at (8.25,2.55) {$T^{(1,1)}$};
\draw[arrow] (9.00, 0.95) -- ++(0.65, 0.65);
\node[ax, anchor=west] at (9.60, 1.65) {$\alpha$};
\node[ax, text=red!70!black, anchor=east] at (7.35,2.10) {$\alpha=0$};
\node[ax, text=blue!70!black, anchor=east] at (8.10,2.87) {$\alpha=1$};
\node[ax, anchor=north] at (7.55,0.25) {$\widetilde{T}\in\mathbb{C}^{2\times4\times4}$};
\node[ax, align=left, anchor=west] at (8.75,3.55) {$\mathbf{I}(0)=(0,0)$\\[-1pt] $\mathbf{I}(1)=(1,1)$};
\end{tikzpicture}
\caption{\textbf{Flat representation of a symmetry-resolved tensor.}
The example shows a two-leg $\mathbb{Z}_2$-symmetric tensor with total charge
$Q=0$ and a uniform intra-sector dimension $d=4$ on each leg.
\textbf{Left:} In the block-sparse representation, each tensor index is written
as a pair $iI$, where $I$ labels the charge sector and $i$ labels the
corresponding intra-sector degree of freedom. Charge conservation permits only
the blocks with $(I_1,I_2)=(0,0)$ and $(1,1)$, while the remaining two sector
combinations are forbidden and therefore absent.
\textbf{Right:} The two symmetry-allowed $4\times4$ blocks are stacked along a
new leading axis to form a single flat tensor $\widetilde{T}$ of shape
$(2,4,4)$. 
The leading index $\alpha \in \{0, 1\}$ enumerates the allowed charge tuples, with $\mathbf{I}(0) = (0,0)$ and $\mathbf{I}(1) = (1,1)$ in this example.
}
\label{fig:flat_repack}
\end{figure}

\paragraph{Flat representation of uniform-shape tensor blocks}

The obstruction identified above originates from the charge dependence of the
dense block shapes. We eliminate this dependence by first imposing a uniform
intra-sector dimension on each tensor leg. Specifically, for leg
$\ell$, we require
\begin{equation}
  D^{(\ell)}_0
  =
  D^{(\ell)}_1
  =
  \cdots
  =
  D^{(\ell)}_{g-1}
  \equiv d_\ell .
  \label{eq:uniform_sector_dimension}
\end{equation}
Under this constraint, the dense shape of a symmetry-allowed block no longer
depends on its charge labels. Every block of a rank-$k$ tensor has the common
shape
\begin{equation}
  d_1 \times \cdots \times d_k .
\end{equation}
The symmetry-allowed blocks can therefore be stacked along an additional
sector axis and represented as a single dense tensor. We refer to this dense
representation of the original block-sparse tensor as its \emph{flat
representation}.

Whilst fixing a uniform initial sector dimension allows the initial configuration projection to be vectorized, for a general symmetry, $G$, subsequent contractions may generate intermediate tensor sectors with sizes that still depend dynamically on the initial configuration, $\sigma$.
This dependence can be eliminated by restricting ourselves to a cyclic symmetry group of any order $g$,
\begin{equation}
  G
  =
  \mathbb{Z}_g .
\end{equation}
The charge sectors on each leg $\ell$ are then labeled by
\begin{equation}
  I_\ell
  \in
  \{0,1,\ldots,g-1\},
\end{equation}
and the charge-conservation rule is evaluated modulo $g$. For a rank-$k$
tensor with total charge $Q\in\mathbb{Z}_g$, the allowed charge-sector tuples
form the set
\begin{equation}
  \mathcal{Q}_T
  =
  \left\{
    (I_1,\ldots,I_k)
    \in
    \mathbb{Z}_g^k
    \;\middle|\;
    \epsilon_1 I_1+\cdots+\epsilon_k I_k
    \equiv
    Q
    \pmod g
  \right\}.
  \label{eq:allowed_charge_tuples}
\end{equation}
The set
$\mathcal{Q}_T$ contains
\begin{equation}
  \lvert \mathcal{Q}_T \rvert
  =
  g^{k-1}
\end{equation}
charge-sector tuples, since any choice of $k-1$ charge labels uniquely
determines the remaining label through the conservation rule modulo $g$. 
Crucially, if the complete set of sectors is present in the initial tensors, then all intermediate contractions will similarly be \emph{sector complete} with shape $g^{k-1}\times d_1\times\cdots\times d_k$, even as the specific $Q$ and $\mathcal{Q}_T$ values depend on configuration.

It is worth noting that this restriction to finite cyclic symmetry
$\mathbb{Z}_g$ does not include the  $\mathrm{U}(1)$ 
symmetry group commonly used in tensor network simulations (e.g. to enforce particle-number symmetry in fermionic simulations). 
However, projection of a tensor network state to a fixed $\mathrm{U}(1)$ charge can easily be enforced during sampling, by restricting the sampled configurations to a
fixed charge sector~\cite{Liu_2017,liu2025fpeps}.

To retain the charge-sector information required for aligning contractions and evaluating fermionic signs, we assign each charge-sector tuple in $\mathcal{Q}_T$ a leading sector index $\alpha \in \{0,\ldots,g^{k-1}-1\}$ and define the corresponding charge table
  \begin{equation}
    \mathbf{I}(\alpha)
    =
    \bigl(
      I_1(\alpha),\ldots,I_k(\alpha)
    \bigr)
    \in
    \mathcal{Q}_T.
    \label{eq:sector_table}
  \end{equation}
In practice, we store $\mathbf{I}$ as a two-dimensional integer array of shape $g^{k-1}\times k$, enabling efficient vectorized sorting, indexing, and manipulation of the sector metadata.
The corresponding symmetry-allowed blocks are then stacked into the flat tensor
\begin{equation}
  \widetilde{T}
  \in
  \mathbb{C}^{g^{k-1}\times d_1\times\cdots\times d_k},
  \label{eq:flat_tensor_shape}
\end{equation}
whose components are defined by
\begin{equation}
\widetilde{T}_{\alpha;\,i_1,\ldots,i_k}
=
T^{\mathbf{I}(\alpha)}_{i_1,\ldots,i_k}
=
T_{i_1 I_1(\alpha),\ldots,i_k I_k(\alpha)}.
\label{eq:flat_tensor}
\end{equation}
The charge table $\mathbf{I}(\alpha)$ and the dense block stack share the leading index $\alpha$, preserving the correspondence between each block and its charge tuple. 
The remaining indices $i_1,\ldots,i_k$ label the dense intra-sector degrees of freedom. 
Figure~\ref{fig:flat_repack} illustrates this packing procedure for a two-leg $\mathbb{Z}_2$-symmetric tensor.

\begin{figure}[t!]
    \centering
    \includegraphics[width=0.8\linewidth]{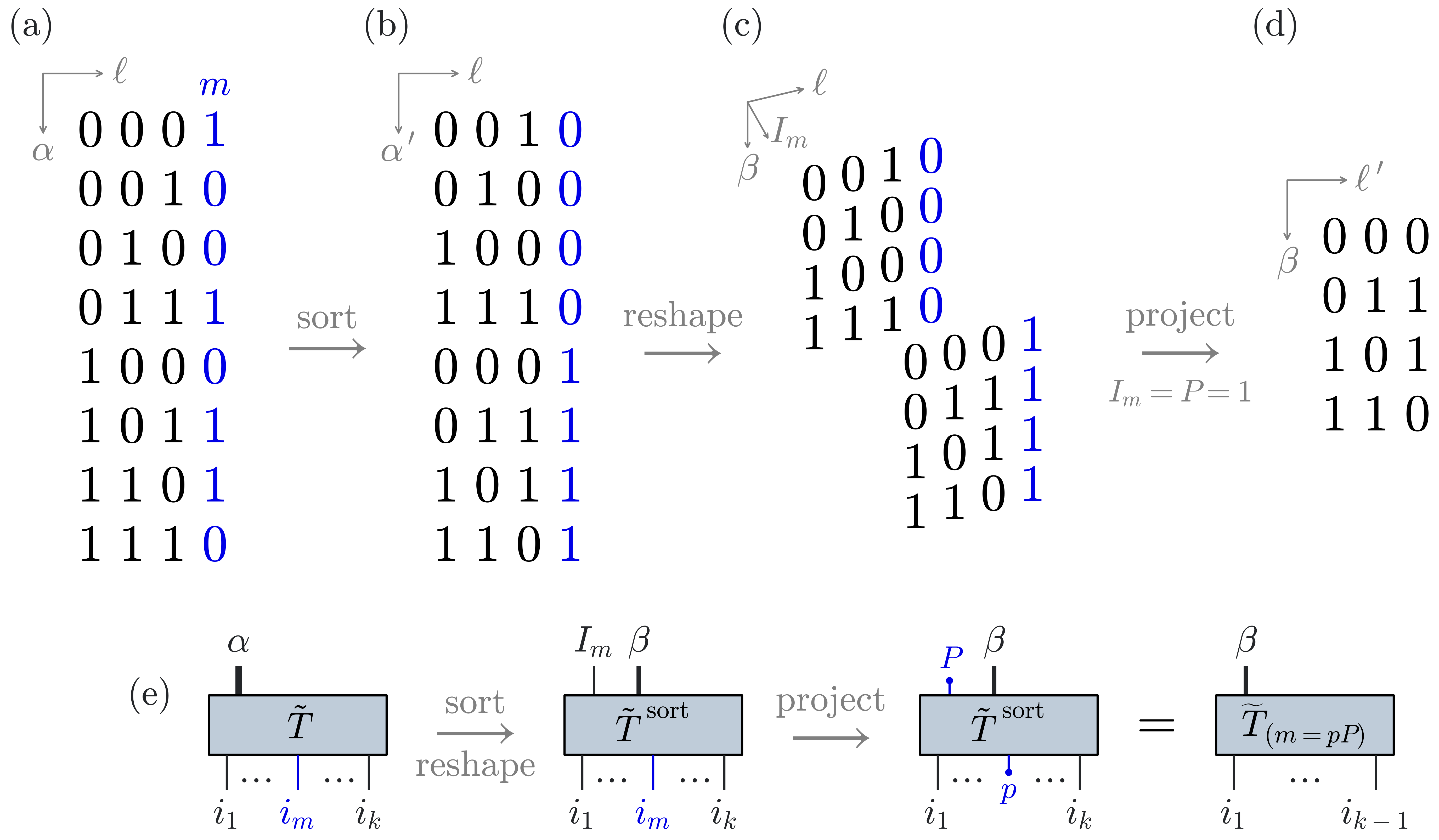}
    \caption{
    An example of sorting, reshaping, and projecting the sector charge table, $\mathbf{I}$, for a $k=4$ rank tensor of $\mathbb{Z}_2$ symmetry and charge $Q=1$, with leg $m$ projected onto charge $P=1$.
    (a) The array is initially indexed by $\alpha$ and $\ell$.
    (b) Sorting by the entries on $m$ yields a new index $\alpha'$.
    (c) This can be reshaped into two sub-indices: $I_m$,  which indexes the charge of leg $m$, and $\beta$, which indexes the remaining charge tuples.
    (d) Finally we project onto the desired value $I_m=P=1$ and drop leg $m$, yielding a new charge table indexed by $\beta$ and $\ell'$.
    (e) The corresponding leading axis transformation and projection of the flat blocks, $\widetilde{T}$, depicted graphically.
    }
    \label{fig:sector-sort-reshape}
\end{figure}

The central advantage of the flat representation is that the charge table can be used to reorder the sectors and corresponding flat blocks along $\alpha$, bringing a tensor into an equivalent representation in which projection, contraction, and other primitives can each be expressed as a single, fixed-shape operation.

Consider projecting the physical leg $m$ of a rank-$k$ flat tensor onto the
symmetry-resolved physical value $i_m=p,~I_m=P$. 
First, the sectors and corresponding dense blocks are sorted by the charge $I_m(\alpha)$ carried by leg $m$, producing a new order $\alpha'$, shown in Fig.~\ref{fig:sector-sort-reshape}(a)-(b). 
Sector completeness ensures that each of the $g$ possible charges occurs exactly
$g^{k-2}$ times, so the sorted sector axis $\alpha'$ can be reshaped into two sub-indices consisting of a new leading charge axis of dimension $g$ for values of $I_m$ followed by a residual sector axis, $\beta$, of dimension $g^{k-2}$ for values of $(I_1,...,I_{m-1},I_{m+1},...,I_k)$:
\begin{equation}
    \widetilde{T}^{\,\mathrm{sort}}
    \in
    \mathbb{C}^{{(g \times g^{k-2}{) \times d_1 \times \cdots \times d_k}}}
    ~.
\end{equation}
This is shown for the corresponding sectors in Fig.~\ref{fig:sector-sort-reshape}(b)-(c).
The projection onto $i_m=p,~I_m=P$ is then a single indexing operation that selects both the symmetry charge $I_m=P$ on the new leading axis  and the intra-sector index $i_m=p$ on leg $m$:
\begin{equation}
    \left(\widetilde{T}_{(m=p P)}\right)_{
    \beta;\,i_1,\ldots,i_{m-1},i_{m+1},\ldots,i_k
    }
    =
    \widetilde{T}^{\,\mathrm{sort}}_{
    P,\beta;\,i_1,\ldots,i_{m-1},p,i_{m+1},\ldots,i_k
}.
\end{equation}
This is shown in Fig.~\ref{fig:sector-sort-reshape}(c)-(d) and (e).
Here $\beta=0,\ldots,g^{k-2}-1$ becomes the new sector index which labels the remaining allowed charge tuples after fixing $I_m=P$. 
The remaining charge tuples automatically satisfy the charge conservation rule:
\begin{equation}
    \epsilon_1 I_1+\cdots+\epsilon_{m-1}I_{m-1}+\epsilon_{m+1}I_{m+1}+\cdots+\epsilon_k I_k
    \equiv
    Q-\epsilon_m P
    \pmod g
  \label{eq:allowed_charge_tuples_after_projection}
\end{equation}
The resulting tensor has shape
$g^{k-2}\times d_1\times\cdots\times d_{m-1}\times
d_{m+1}\times\cdots\times d_k$, independent of the selected physical index value.

For a site tensor $\widetilde{T}^{[i]}$, write the local physical state as $\sigma_i=p P$. We denote the resulting projected tensor by
\begin{equation}
  \widetilde{A}^{[i]}(\sigma_i)
  \equiv
  \widetilde{T}^{[i]}_{(\mathrm{phys}=\sigma_i)}
  \in
  \mathbb{C}^{
    g^{k-2}
    \times
    d_1
    \times\cdots\times
    d_{m-1}
    \times
    d_{m+1}
    \times\cdots\times
    d_k
  }.
  \label{eq:fermionic_projected_site_tensor}
\end{equation}
Because $\widetilde{A}^{[i]}(\sigma_i)$ has the same shape for every local
state $\sigma_i$, the projected tensors for a batch of configurations can be
stacked along a leading batch dimension:
\begin{equation}
  \widetilde{A}^{[i]}_B
  =
  \operatorname{stack}_{a=1}^{B}
  \widetilde{A}^{[i]}
  \!\left(
    \sigma_i^{(a)}
  \right)
  \in
  \mathbb{C}^{
    B
    \times
    g^{k-2}
    \times
    d_1
    \times\cdots\times
    d_{m-1}
    \times
    d_{m+1}
    \times\cdots\times
    d_k
  }.
  \label{eq:fermionic_batched_projected_tensor}
\end{equation}
This allows the initial configuration projection step in an amplitude computation to be vectorized.

\begin{figure}[t!]
    \centering
    \includegraphics[width=0.8\linewidth]{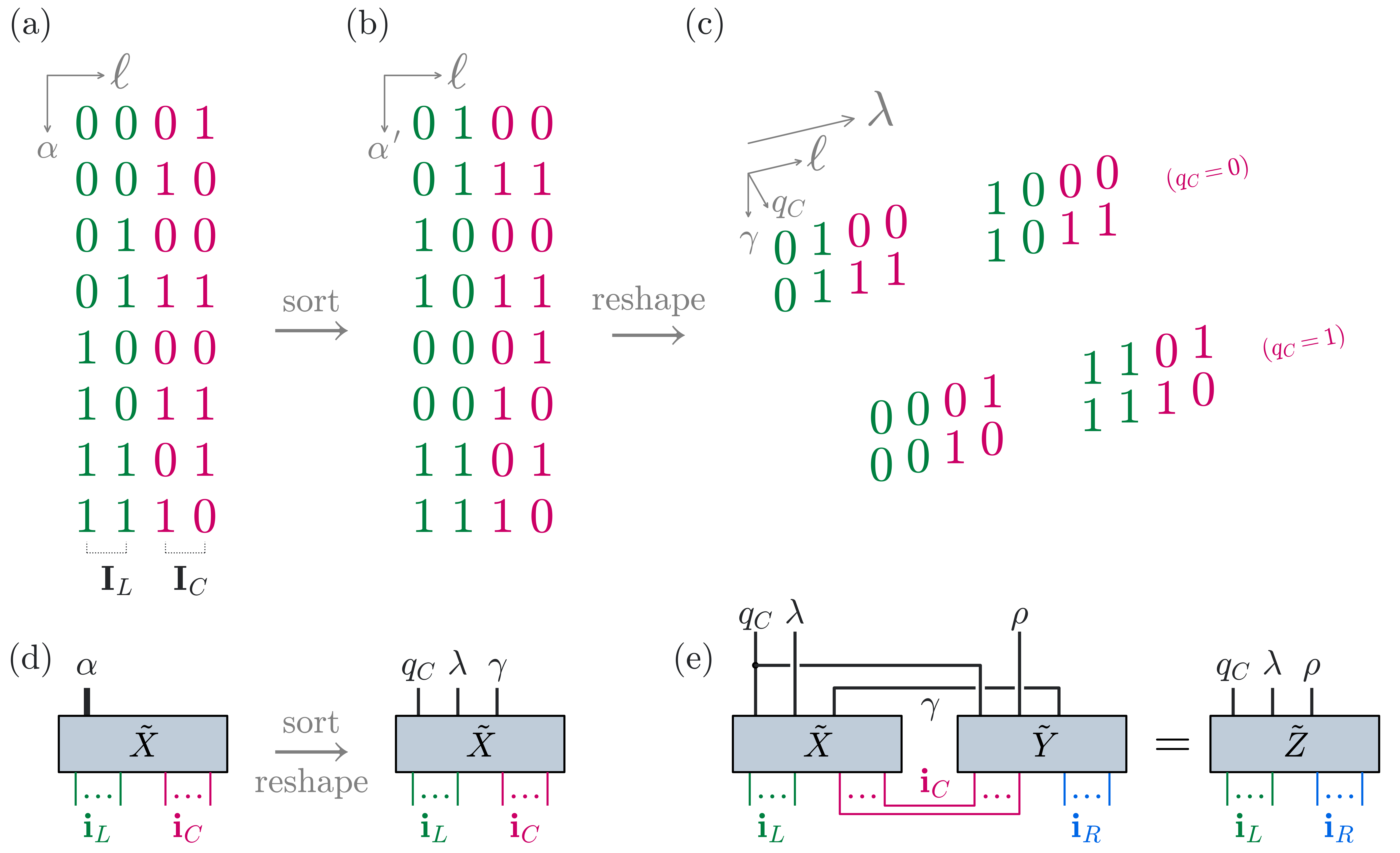}
    \caption{
    An example of sorting and reshaping the sector charge table $\mathbf{I}$ of a flat tensor $\widetilde{X}$ with $\mathbb{Z}_2$ symmetry and charge $Q=1$ to prepare for contraction.
    (a) The array is initially indexed by sector index $\alpha$ and leg index $\ell$. $\mathbf{I}_L$ denotes the charge columns of legs to be kept, $\mathbf{I}_C$ those to be contracted away.
    (b) Sorting the rows lexicographically by $\bigl(q_C,\mathbf{I}_L,\mathbf{I}_C \bigr)$ yields a new sector index $\alpha'$.
    (c) This can be reshaped into three sub-indices: $q_c$, which indexes the \emph{combined} contracted charge (row-sum of $\ell \in C$);
    $\lambda$, which indexes the possible charge sectors on the kept legs;
    and $\gamma$, which indexes the remaining charge sectors on the contracted legs.
    (d) The corresponding leading axis transformation on the flat blocks, $\widetilde{X}$, depicted graphically.
    (e) Once in such form, two arrays can be directly contracted using Eq.~\eqref{eq:flat_direct_contraction}, shown here graphically.
    }
    \label{fig:sector-sort-reshape-contract}
\end{figure}

\paragraph{Direct flat contraction.}

The flat representation also allows tensor contraction to be performed as a single dense operation after suitably sorting and reshaping the sector axes $\alpha$ of the two tensors.
Consider contracting two sector-complete flat tensors $\widetilde{X}$ and $\widetilde{Y}$ of ranks $k_X$ and $k_Y$ with finite cyclic symmetry $\mathbb{Z}_g$ over $n$ paired legs.
Write their kept legs as the ordered tuples $L$ and $R$, and their contracted legs as the ordered tuples $C=(c_1,\ldots,c_n)$ and $C'=(c'_1,\ldots,c'_n)$, where $c_j$ is contracted with $c'_j$.
For a sector index $\alpha$ of the charge table and an ordered set of legs $S=(s_1,\ldots,s_m)$, define its charge tuple by $\mathbf{I}_S(\alpha)=(I_{s_1}(\alpha),\ldots,I_{s_m}(\alpha))$ and its effective charge relative to the orientation of its first leg by
\begin{equation}
    q_S(\alpha)=\epsilon_{s_1}\sum_{j=1}^{m}\epsilon_{s_j}I_{s_j}(\alpha)\pmod g.
    \label{eq:effective_charge}
\end{equation}
We suppress the dependence on the sector index below.
We use $q_C$ to denote the effective charge of the contracted legs.
For every $j$, the paired legs $c_j$ and $c'_j$ carry the same charge label and opposite orientations, thus $\mathbf{I}_C=\mathbf{I}_{C'}$ and $q_C=q_{C'}$ and the charge tables of $\tilde{X}$ and $\tilde{Y}$ can be sorted independently, with the elements remaining appropriately aligned for the contraction.

The charge table of $\widetilde{X}$ is sorted lexicographically by
\begin{equation}
  \bigl(
    q_C,\mathbf{I}_L,\mathbf{I}_C
  \bigr),
  \label{eq:Xtuples}
\end{equation}
while that of $\widetilde{Y}$ is sorted by
\begin{equation}
  \bigl(
    q_C,\mathbf{I}_R,\mathbf{I}_{C'}
  \bigr),
  \label{eq:Ytuples}
\end{equation}
with the earlier entries serving as the more significant sort keys.
An example is shown in Fig.~\ref{fig:sector-sort-reshape-contract}(a)-(b).
After sorting, the first key groups blocks which have the same combined charges of the contracted legs.
The second key further groups blocks by their eventual output sector.
The third key finally aligns charge-tuples of the contracted legs such that they are summed pairwise in inner-product fashion.


Sector completeness then allows the sorted sector axes to be reshaped as
\begin{equation}
  \alpha_X
  \longmapsto
  (q_C,\lambda,\gamma),
  \qquad
  \alpha_Y
  \longmapsto
  (q_C,\rho,\gamma).
\end{equation}
This is shown in Fig.~\ref{fig:sector-sort-reshape-contract}(b)-(c) and (d).
Here $q_C\in\mathbb{Z}_g$, $\lambda=0,\ldots,g^{|L|-1}-1$, $\rho=0,\ldots,g^{|R|-1}-1$, and $\gamma=0,\ldots,g^{n-1}-1$.
Likewise, let $\mathbf{i}_S=(i_\ell)_{\ell\in S}$ denote the ordered intra-sector multi-index on $S$.
For the paired contracted legs, we use the common multi-index $\mathbf{i}_C$ on both tensors.
Writing $(d_1^{(X)},\ldots,d_{k_X}^{(X)})$ and $(d_1^{(Y)},\ldots,d_{k_Y}^{(Y)})$ for the dense block dimensions of the two tensors, the reshaped blocks can be written as
\begin{equation}
  \begin{aligned}
    \widetilde{X}_{q_C,\lambda,\gamma;\,\mathbf{i}_L,\mathbf{i}_C}
    &\in
    \mathbb{C}^{g\times g^{|L|-1}\times g^{n-1}\times d_1^{(X)}\times\cdots\times d_{k_X}^{(X)}},
    \\
    \widetilde{Y}_{q_C,\rho,\gamma;\,\mathbf{i}_C,\mathbf{i}_R}
    &\in
    \mathbb{C}^{g\times g^{|R|-1}\times g^{n-1}\times d_1^{(Y)}\times\cdots\times d_{k_Y}^{(Y)}}.
  \end{aligned}
\end{equation}

The complete block-sparse contraction can then be evaluated as the single dense contraction
\begin{equation}
  \widetilde{Z}_{q_C,\lambda,\rho;\,\mathbf{i}_L,\mathbf{i}_R}
  =
  \sum_{\gamma,\mathbf{i}_C}
  \widetilde{X}_{q_C,\lambda,\gamma;\,\mathbf{i}_L,\mathbf{i}_C}
  \widetilde{Y}_{q_C,\rho,\gamma;\,\mathbf{i}_C,\mathbf{i}_R}.
  \label{eq:flat_direct_contraction}
\end{equation}
This is depicted graphically in Fig.~\ref{fig:sector-sort-reshape-contract}(e).
The labels $q_C$, $\lambda$, and $\rho$ are flattened to form the sector axis $\alpha_Z$ of the result, which contains $g^{|L|+|R|-1}$ uniform-shape blocks.
Thus a flat tensor is directly produced.
All array shapes and the sequence of operations are configuration-independent, so the whole process is readily vectorized.
This operation is analogous to the reduced-form contraction described in Ref.~\cite{SciPostPhysCodeb.10}. We provide a concrete example of the direct flat contraction in Appendix~\ref{app:flat_contraction_example}.

We also note here that leg fusion, needed to put tensors into matrix form for SVD and other linear-algebra decompositions, uses a very similar sort-and-reshape procedure, except that the analogous subsector label $\gamma$ is \emph{combined} with the intra-sector index of the fused leg rather than summed over.

\paragraph{Fermionic tensor networks and branchless sign tracking.}

Fermionic tensor networks are
constructed from fermionic tensors that obey a $\mathbb{Z}_2$-graded algebra which encodes fermion parity and anticommutation relations; we refer the
reader to Ref.~\cite{gao2024fermionic,Mortier_2025} for more technical details. Consequently,
fermionic tensors are necessarily symmetric under the $\mathbb{Z}_2$
fermion-parity symmetry. Computations with fermionic tensor networks can therefore naturally use the symmetry tensor network operations defined above. 

However, fermionic tensors also include an ordering of the tensor network legs~\cite{gao2024fermionic}, and additional  parity-dependent signs appear when the tensor legs are reordered. In the flat representation, the charge tuple associated with block $\alpha$ is stored in the charge table $\mathbf{I}(\alpha)$. 
The sign produced by exchanging legs $r$ and $s$ is therefore
\begin{equation}
\eta_{rs}(\alpha)
=
(-1)^{p(I_r(\alpha))p(I_s(\alpha))},
\label{eq:flat_swap_sign}
\end{equation}
where $p(I) \in \{0,1\}$ is the fermion parity of charge $I$.
A general permutation composed of such swaps produces a vector of signs, $\phi(\alpha)$, computed solely from the charge table.
This can be retained separately to accumulate signs from other operations, and multiplied into the data blocks when necessary, for example prior to contraction, via a broadcasted multiply $T'_{\alpha,i_1,\ldots,i_k} = \phi(\alpha) T_{\alpha,i_1,\ldots,i_k}$.


Signs associated with general permutations, fusions, decompositions, and contractions are all obtained from this elementary swap operation. 
Although their values may depend on the current charge sectors, they are computed through fixed-shape algebraic operations on the charge metadata and applied element-wise to the flat blocks, without inspecting their numerical entries or introducing configuration-dependent control flow. 
Together with the uniform flat layout and cyclic symmetry, this branchless sign tracking yields a static computational graph suitable for vectorized fermionic tensor-network amplitude evaluation.

\section{Numerical results for fermionic TN-VMC}
\label{sec:results}

The flat symmetric tensor formalism developed above provides the
configuration-independent amplitude-evaluation primitive required to realize the GPU-adapted symmetric TN-VMC workflow illustrated in
Fig.~\ref{fig:vmc_parallel_schemes}(c). In this section, we use fermionic TN as the example of a symmetric TN, and examine how this
primitive performs in practical VMC calculations for the two-dimensional
Fermi--Hubbard model.

We first consider two representative variational
functions based on fermionic tensor networks with $\mathbb{Z}_2$ symmetry: TNF-fPEPS, whose amplitudes are defined by a fixed approximate
contraction procedure, and standard fPEPS, for which efficient VMC requires
boundary-MPS environment reuse. These two cases test, respectively, direct
vectorization of a fixed amplitude graph and vectorization of a finite set of
geometry-dependent graphs. We then examine multi-GPU scaling and
production-scale wall times, test the portability of the vectorized fTN
amplitude construction across independent tensor backends, and demonstrate its
application to a deep NN-fPEPS. In our calculations, the variational parameters and the wavefunction
amplitudes are restricted to be real-valued, such that
$\Psi_\theta(\sigma)\in\mathbb{R}$ for all physical configurations $\sigma$.
The variational parameters are represented in single precision.

All CPU runs use Intel Xeon Platinum 8352Y (Ice Lake) processors, and all GPU
runs use NVIDIA A100 GPUs with 80~GB of memory. 

\subsection{Single GPU wall-time reduction}
\label{sec:single_rank_walltime}

We first analyze the single GPU (single MPI rank) performance gain obtained from vectorized fTN
amplitude evaluation. The GPU uses a batched Markov chain with $B$ configurations. To reflect a baseline that uses a standard implementation of tensor networks, we compare against a single CPU core executing a single Markov chain.

\begin{figure}[!h]
\centering
\begin{subfigure}[!htbp]{0.48\textwidth}
  \centering
  \includegraphics[width=\linewidth]{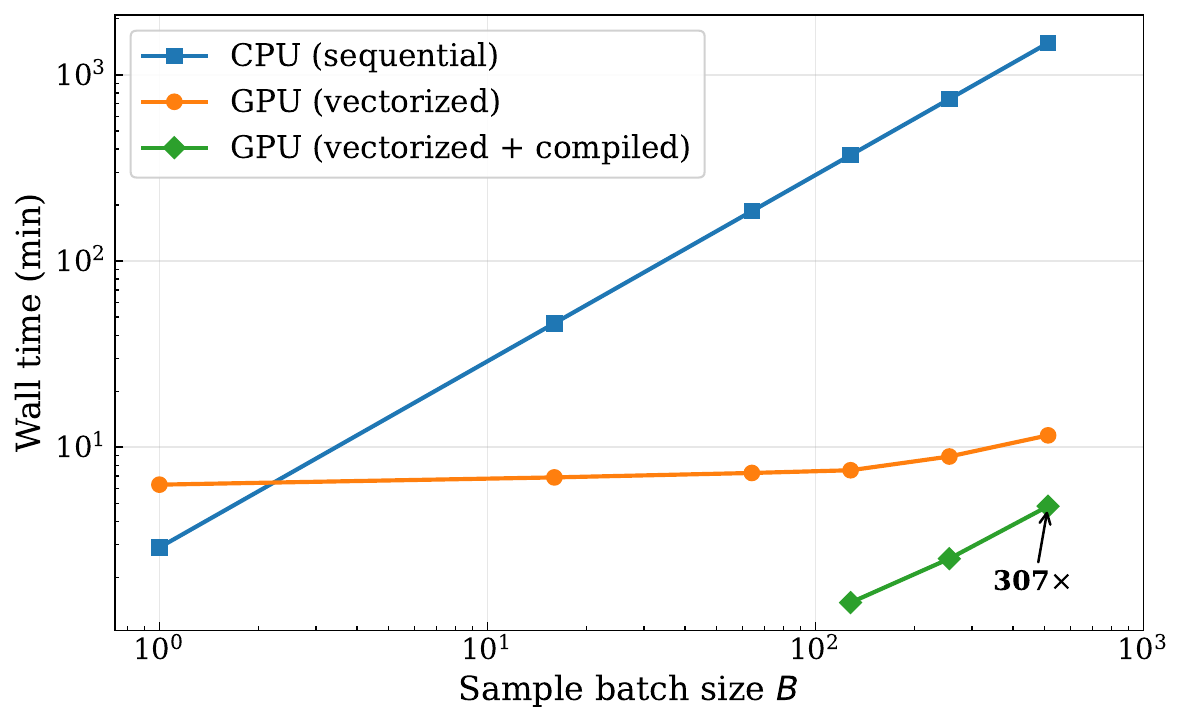}
  \caption{TNF-fPEPS on an $8\times 8$ lattice.}
  \label{fig:tnf_speedup}
\end{subfigure}
\hfill
\begin{subfigure}[!htbp]{0.48\textwidth}
  \centering
  \includegraphics[width=\linewidth]{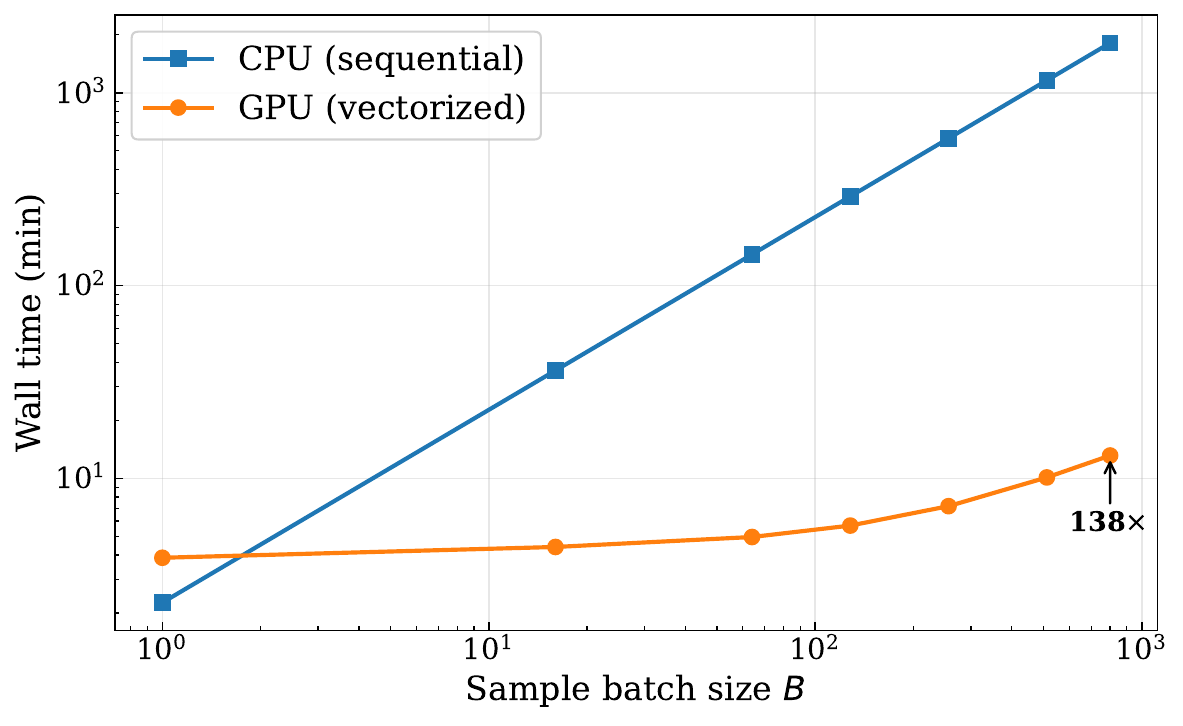}
  \caption{Standard fPEPS on a $16\times 8$ lattice.}
  \label{fig:bmps_reuse_speedup}
\end{subfigure}
\caption{\textbf{Single-rank VMC step time as a function of the sample batch size $B$.}
(a) TNF-fPEPS with bond dimension $D=12$ and boundary-MPS bond dimension
$\chi=12$. (b) Standard fPEPS with bond dimension $D=8$ and boundary-MPS bond
dimension $\chi=32$, using boundary-MPS environment reuse. CPU timings use one
MPI rank with a single CPU core and sequential amplitude
evaluation. GPU timings use one MPI rank controlling one GPU
and batched amplitude evaluation with batch size $B$.}
\label{fig:single_rank_speedup}
\end{figure}

\subsubsection{Tensor network functions}
\label{sec:tnf}

We begin with TNF-fPEPS, which provides the most direct realization of the
vectorized fTN amplitude construction. In a TNF, the amplitude is defined by a fixed prescribed 
tensor-network contraction procedure, including the truncations used during
the contraction. The resulting function therefore need not coincide with the
exact contraction of an underlying tensor-network state, but it must define a
unique amplitude for every input configuration.

We choose as the underlying state an open-boundary fPEPS with bond dimension $D$.
For a physical configuration $\sigma$, the single-layer
amplitude network is approximately contracted using the boundary-MPS
method~\cite{verstraete2004renormalization,corboz2016variational} with bond
dimension $\chi$. We fix the row-absorption direction and contraction
ordering for all configurations, so that the contraction procedure defines a
single configuration-independent computation which gives the amplitude function $\sigma\to \Psi(\sigma)$. The flat fermionic
representation further ensures that the projected tensor layouts and
fermionic sign operations within this graph are independent of $\sigma$.
The resulting TNF-fPEPS amplitudes can therefore be evaluated directly as one
dense vectorized batch.

Figure~\ref{fig:tnf_speedup} shows the wall time per VMC step for the
2D Fermi--Hubbard model on an $8\times8$ lattice, using a TNF-fPEPS
with bond dimension $D=12$ and boundary-MPS bond dimension $\chi=12$. 
Here the total number of VMC samples generated per step is equal to the sample batch size $B$. The GPU workflow timings are measured for different batch sizes
$B$, both with and without GPU just-in-time (JIT) compilation. 
The CPU reference uses a single Markov chain to collect the same number of $B$ samples. 
At $B=512$, the compiled GPU workflow is approximately $307\times$ faster than the
CPU reference. Even without compilation, the batched GPU implementation already
provides more than an order-of-magnitude reduction in wall time.

\subsubsection{Standard fPEPS with boundary-MPS reuse}
\label{sec:standard_fpeps_bmps_reuse}

We next consider standard fPEPS where the goal is to contract the TN amplitude accurately (approximating the exact contraction limit) rather than the amplitude being intrinsically defined through approximate contraction as for TNF. This is the variational ansatz used in standard fPEPS-VMC calculations~\cite{liu2025fpeps}.
A standard fPEPS-VMC calculation reuses
intermediate boundary-MPS environments during sampling and local-energy
evaluation, reducing the cost of a VMC step from $O(N^2)$ to $O(N)$ in the
number of lattice sites $N=L_xL_y$~\cite{Liu_2017,Liu2021accuratepeps}.
A meaningful GPU benchmark must therefore employ this environment-reuse
strategy.

The flat fermionic representation removes the configuration dependence of the
local tensor layouts and fermionic sign operations, as in the TNF-fPEPS case.
Environment reuse introduces an additional algorithmic structure: the
amplitudes required during a VMC step are assembled from different cached
boundary environments and therefore do not all share one global contraction
graph. They instead form a finite set of geometry classes. Within each class,
the residual amplitude networks have common tensor shapes and a shared
contraction computational graph and can be evaluated as a vectorized batch. The construction
is detailed in Appendix~\ref{app:bmps_reuse}.

Figure~\ref{fig:bmps_reuse_speedup} reports the wall time per VMC step (the VMC sample size is the sample batch size $B$) for
standard fPEPS on a $16\times8$ lattice with bond dimension $D=8$ and
boundary-MPS bond dimension $\chi=32$. ($\chi$ here is large enough to
remove the contraction error in amplitudes within the variational accuracy of the fPEPS~\cite{liu2025fpeps}).
At $B=800$, the GPU pipeline is
approximately $138\times$ faster than the single core CPU reference. The speedup is smaller
than in the TNF-fPEPS benchmark, primarily because we did not use JIT compilation, which for large lattices can require long compile times due to the number of different computational graphs (coming from the many boundary environments).

\begin{figure}[!ht]
\centering
\includegraphics[width=0.75\columnwidth]{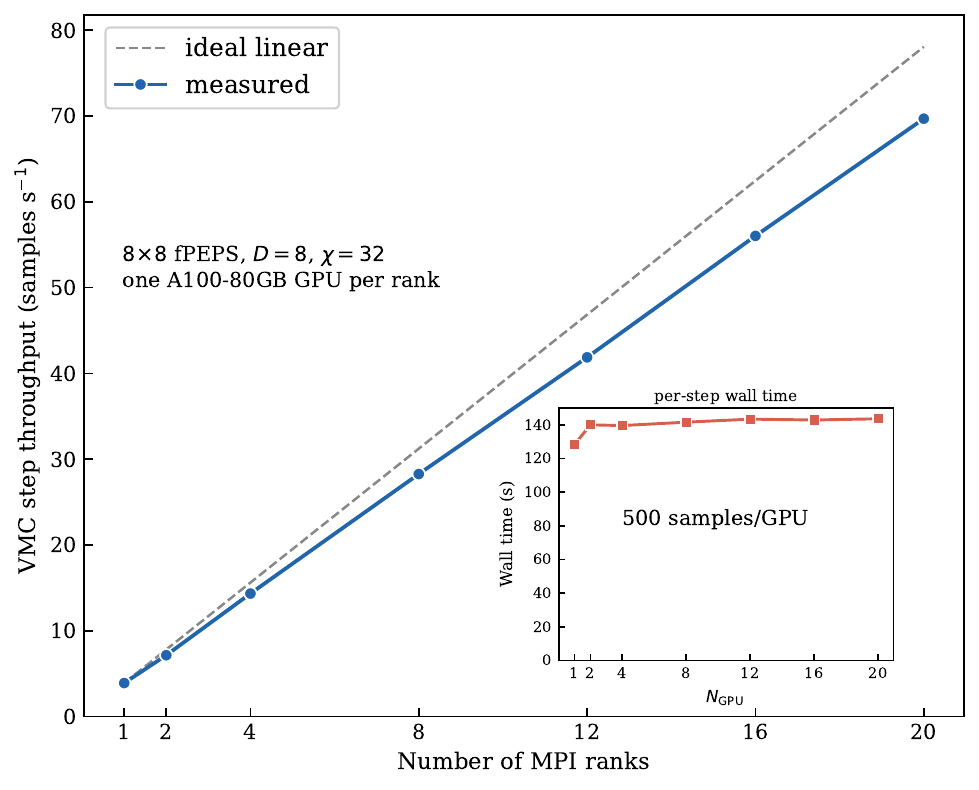}
\caption{\textbf{Multi-GPU throughput scaling of batched TN-VMC.}
Standard fPEPS-VMC on an $8\times8$ Fermi--Hubbard lattice with bond dimension
$D=8$ and boundary-MPS bond dimension $\chi=32$. Each MPI rank controls one
GPU and carries a fixed batch of $B=500$ walkers. The main
panel shows the total number of Monte Carlo samples processed per second of
VMC-step wall time as a function of the number of MPI ranks/GPUs; the dashed
line indicates ideal linear throughput scaling. The inset shows the VMC-step
wall time remains constant as the number of GPUs is increased.}
\label{fig:gpu_multi_scaling}
\end{figure}

\subsection{Multi-GPU throughput scaling}
\label{sec:multi_gpu_scaling}

We next examine the performance of the batched TN-VMC GPU implementation as a function of the number of GPUs.
Here
we keep the per-GPU sample batch size fixed and increase the total number of samples
by adding more GPUs. This setting corresponds to the typical production use
case where larger Monte Carlo sample sizes are used to reduce
statistical uncertainty. For the calculations, we use a multi-GPU parallelization using MPI. 

Figure~\ref{fig:gpu_multi_scaling} reports the throughput of a standard
fPEPS-VMC calculation on an $8\times8$ Fermi--Hubbard lattice with bond
dimension $D=8$ and boundary-MPS bond dimension $\chi=32$. Each MPI process
controls one GPU and carries a fixed batch of $B=500$ walkers.
The number of GPUs, which is equivalent to the total number of MPI ranks, is varied from 1 to
20. The throughput is measured as the number of Monte Carlo samples processed
per second of VMC-step wall time.

The total throughput increases nearly linearly with the number of GPUs. The
inset shows the corresponding wall time per VMC step as $N_{\mathrm{GPU}}$ is
increased at fixed per-GPU batch size; this time remains nearly constant over
the measured range. These results indicate that the collective-communication
overhead is small compared with the local batched tensor contractions in this
regime. Thus, for the system sizes and batch sizes considered here, larger VMC
sample sizes can be efficiently handled by adding more GPUs, without
substantially increasing the wall time per VMC step.
We note that this near-ideal scaling relies on each GPU
carrying a batch size large enough to saturate the device; for a \emph{fixed}
total sample budget, distributing the samples over more GPUs eventually
shrinks the per-GPU batch size into the underutilization regime characterized in
Sec.~\ref{sec:single_rank_walltime}, so the useful number of GPUs is bounded
by the per-GPU batch size required for efficient device utilization.

\subsection{End-to-end speedup in production-scale VMC runs}
\label{sec:hpc_matched}

In the following, we demonstrate the end-to-end VMC wall-time 
reduction using multiple GPUs in representative large-scale calculations. 
We consider VMC runs using
standard fPEPS on a $16\times8$ Hubbard
lattice and TNF-fPEPS on an $8\times8$ Hubbard lattice, each with $20{,}000$
Monte Carlo samples per VMC step. 
We use two hardware configurations that reflect a possible real-life computational cluster setup: 20 GPUs versus 800 CPU cores (consistent with a 40:1 CPU core to GPU ratio hardware build).  
The CPU allocation used
$800$ MPI processes, i.e. $1$ CPU 
core per process.
The GPU allocation used 20 MPI processes, i.e. 1 GPU per process. 

Table~\ref{tab:hpc_matched} reports the resulting wall time per VMC step. 
For
standard fPEPS, the GPU run completes one VMC step in $0.33$~h, compared with
$0.94$~h on the CPU allocation. This is a $3\times$
reduction, or equivalently, a $120\times$ speedup over the single CPU core. For TNF-fPEPS, the GPU run completes one VMC step in $0.16$~h,
compared with $1.21$~h on the CPU allocation, corresponding to an approximately
$8\times$ speedup, or $\sim 300\times$ the single CPU core performance. These results show that the multi-GPU throughput on a typical cluster translates into a considerable
speedup in production-scale TN-VMC calculations.

\begin{table}[!htbp]
\centering
\begin{tabular}{lccc}
\toprule
Ansatz / system & CPU (800 cores) & GPU (20) & Speedup \\
\midrule
Standard fPEPS, $16{\times}8$, $D=8$, $\chi=32$
  & $0.94$~h & $0.33$~h & $\sim 3\times$   \\
TNF-fPEPS, $8{\times}8$, $D=12$, $\chi=12$
  & $1.21$~h  & $0.16$~h & $\sim 8\times$  \\
\bottomrule
\end{tabular}
\caption{\textbf{Wall time per VMC step in production-scale runs.}
Each VMC step uses $20{,}000$ Monte Carlo samples. The CPU allocation uses
$800$ MPI ranks with $1$ CPU core per rank, for $800$
CPU cores in total. The GPU allocation uses $20$ GPUs.}
\label{tab:hpc_matched}
\end{table}

\subsection{Backend portability of vectorized fTN amplitudes}
\label{sec:backend_independence}

The vectorized tensor network amplitude formalism developed in
Sec.~\ref{sec:vectorization} is not tied to a specific tensor library or
computational backend. 


\begin{figure}[!h]
\centering
\includegraphics[width=0.75\columnwidth]{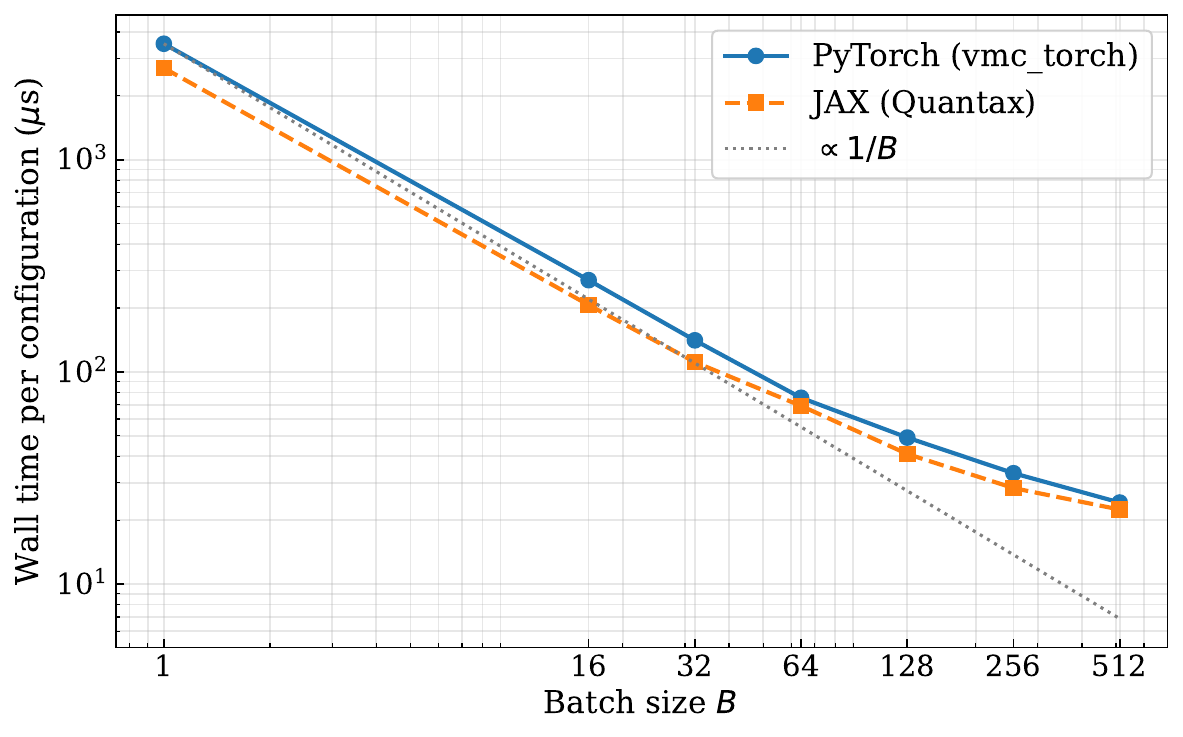}
\caption{\textbf{Backend independence of vectorized tensor network amplitude
evaluation.}
Amortized amplitude-evaluation wall time per configuration for a
$16\times4$ TNF-fPEPS with bond dimension $D=4$ and boundary MPS bond dimension $\chi=\infty$ (exact contraction). 
The two curves correspond to independent
implementations using \texttt{vmc\_torch} and \texttt{Quantax}.
The calculation is performed on a single GPU. }
\label{fig:backend_independence}
\end{figure}

We demonstrate this backend portability using two independent VMC implementations:
\texttt{vmc\_torch}~\cite{Du_vmc_torch_Flexible_Variational}, based on PyTorch~\cite{torch2019github}, and
\texttt{Quantax}~\cite{quantax}, based on JAX~\cite{jax2018github}. We measured the wall-clock time for batched amplitude
evaluation of a $16\times4$ TNF-fPEPS with bond dimension $D=4$, using exact
contraction, over a range of batch sizes $B$ on a single GPU, as shown in
Fig.~\ref{fig:backend_independence}.

The PyTorch and JAX implementations exhibit comparable wall-clock times and
similar dependence on the batch size. In both cases, the amortized cost per
configuration decreases approximately as $1/B$ at small and intermediate batch
sizes and eventually saturates once the available GPU parallelism is fully
utilized. The close agreement between the two independent implementations
indicates that the observed batching behavior is an intrinsic property of the
vectorized fTN amplitude construction rather than a backend-specific feature.

\subsection{Application: NN-fTNS with deep neural networks}
\label{sec:nnfpeps_application}

We finally apply the vectorized fTN amplitude algorithm to a VMC optimization
of a NN-fTNS~\cite{du2024neuralized}.
An NN-fTNS augments an fTNS with a neural network that maps the physical
configuration $\sigma$ to configuration-dependent tensor parameters.
Its amplitude is then obtained by contracting the resulting amplitude tensor network.
This hybrid construction combines the
structured fermionic many-body representation of an fTNS with the additional
expressivity of a neural network.

\begin{figure}[!ht]
\centering
\includegraphics[width=0.85\columnwidth]{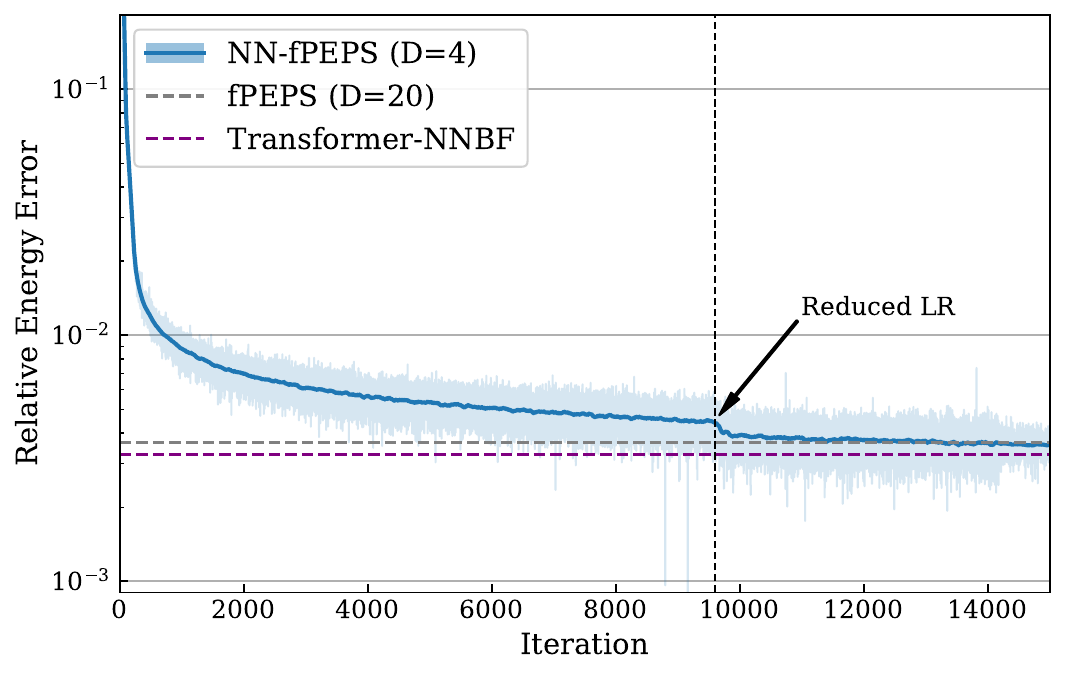}
\caption{\textbf{Ground-state VMC optimization of a deep NN-fPEPS for a $16\times 4$ Hubbard model.}
The Hamiltonian is defined on a square lattice with open boundary conditions with the parameters $U/t=8$
and hole doping $\delta=1/8$.
The relative energy error is calculated with respect to the extrapolated exact DMRG ground state energy~\cite{liu2025fpeps}. 
The NN-fPEPS has bond dimension $D=4$ and is coupled to a deep neural network with $16$ hidden layers. 
We see that after $\sim 14000$ optimization steps, the NN-fPEPS energy surpasses the accuracy of the prior reported 
fPEPS~\cite{liu2025fpeps} energy that uses a substantially larger bond dimension ($D=20$). The energy is also competitive with the recent neural quantum state result based on a Transformer NN-backflow architecture~\cite{gu2025solvinghubbardmodelneural}. 
The NN-fPEPS optimization is performed using \texttt{vmc\_torch} on
$8$ GPUs with a total Monte Carlo sample size of $2048$
per step. We reduce the optimization learning rate (LR) at step $9600$ for faster convergence.}
\label{fig:nnfpeps_energy}
\end{figure}

Previous NN-fTNS calculations relied on conventional CPU-based TN-VMC
workflows and were limited to relatively shallow neural networks because of
the cost of sequential amplitude evaluation~\cite{du2024neuralized}. In the
present workflow, both the neural-network evaluation and the subsequent flat
fermionic tensor-network contraction are performed in batches on GPUs. This
fully vectorized GPU implementation makes the efficient optimization of
NN-fTNS with substantially deeper neural networks computationally practical.

As an illustrative example, Fig.~\ref{fig:nnfpeps_energy} shows the optimization
of a neuralized fPEPS (NN-fPEPS) incorporating a deep convolutional neural network with
$16$ hidden layers. We consider the two-dimensional Fermi--Hubbard model at
$U/t=8$ and hole doping $\delta=1/8$ on a $16\times4$ square lattice with open
boundary conditions. Despite its small tensor-network bond dimension $D=4$,
the deep NN-fPEPS achieves a ground-state energy comparable to those obtained
with a conventional fPEPS of substantially larger bond dimension
$D=20$~\cite{liu2025fpeps}, as well as state-of-the-art fermionic neural quantum
states~\cite{gu2025solvinghubbardmodelneural}. We present this result as a
representative demonstration of the capabilities enabled by the GPU-accelerated
workflow; a systematic investigation of deep NN-fTNS is deferred to future
work.

\section{Software implementation}
\label{sec:vmc_torch}

The methods developed in this work were implemented in a software stack
combining \texttt{quimb}~\cite{gray2018quimb},
\texttt{symmray}~\cite{gao2024fermionic,symmray}, and a VMC backend (either 
\texttt{vmc\_torch}~\cite{Du_vmc_torch_Flexible_Variational}, used in most of the calculations, or \texttt{Quantax}~\cite{quantax}, used also in Sec.~\ref{sec:backend_independence}).
This structure separates the tensor-network
representation (\texttt{quimb}), fermionic tensor algebra (\texttt{symmray}), and VMC workflow (\texttt{vmc\_torch} and \texttt{Quantax}). Note that while the batched support for symmetric tensors in \texttt{symmray} is limited to finite cyclic symmetry groups, \texttt{symmray} itself can handle general abelian symmetries, such as $\mathrm{U}(1)$ symmetry, with charge dependent sector dimensions. This functionality is also compatible with the CPU version of \texttt{vmc\_torch}, allowing for (non-batched) CPU-based VMC calculations with general abelian symmetric tensors.
\section{Summary and outlook}
\label{sec:summary}

We have developed a vectorized formulation of symmetric tensor network
operations. By combining a flat symmetric tensor formalism with 
a uniform-shape tensor layout, with branchless fermionic sign tracking, we achieved vectorized amplitude evaluation for fTNs
with finite cyclic symmetries, such as $\mathbb{Z}_g$.
Based on this vectorized formulation, we implemented a GPU-adapted TN-VMC formulation, which we benchmarked
on the two-dimensional Fermi--Hubbard model using both TNF-fPEPS and
standard fPEPS. The implementation exhibited
near-linear multi-GPU throughput scaling, and achieved up to $300 \times$ end-to-end wall-time speedup over a single CPU core
for both TNF-fPEPS and standard
fPEPS.

This level of GPU acceleration will enable substantially larger TNS and NN-fTNS ans\"atze to be used. As an example, we
trained an NN-fPEPS with a deep neural network and obtained an energy accuracy
competitive with a much larger bond dimension fPEPS and a state-of-the-art fermionic NQS. 

Beyond VMC for quantum many-body ground states, vectorized tensor-network computation
may also be
useful in applications where many tensor networks of the same structure must be
contracted repeatedly. One example is tensor network Monte Carlo, where vectorized tensor-network
computation combined with GPU may accelerate parallel configuration sampling for statistical-physics
models such as spin
glasses~\cite{chen2025tensornetworkmarkovchain,chen2025batchtnmcefficientsamplingtwodimensional}
and continuous-space systems~\cite{park2026statisticalmechanicscontinuousspace}.

\section*{Acknowledgments}
S.D. acknowledges useful discussions with Jielun Chen during
the preparation of this manuscript. This work was supported by the US Department of Energy, Office of Science, via grant no. DE-SC0019374. We used the computing resources of the National Energy Research Scientific Computing Center (NERSC), a Department of Energy User Facility, using NERSC award ERCAP-38682.

\begin{appendix}
\section{Example: contraction of two rank-five tensors with $\mathbb{Z}_2$ symmetry in flat representation}
\label{app:flat_contraction_example}
Consider two rank-five block-sparse tensors of $\mathbb{Z}_2$ symmetry and $\mathbb{Z}_2$ charge $Q=0$, written in the flat representation,
\begin{equation}
  \widetilde{X}_{a b c\,u v},
  \qquad
  \widetilde{Y}_{u v\,d e f},
\end{equation}
The two shared indices 
$(u, v)$ are contracted to obtain a rank-six $\mathbb{Z}_2$
tensor $\widetilde{Z}_{abc,def}$ with $\mathbb{Z}_2$ charge $Q=0$. The goal is to perform the block-sparse tensor contraction as a dense tensor contraction in the flat representation, and directly obtain the output tensor $\widetilde{Z}$ in its flat representation. 

Both $\widetilde{X}$ and $\widetilde{Y}$ contain $2^{5-1}=16$
symmetry-allowed dense tensor blocks of equal shape. In the flat representation, we stack them into a dense tensor with leading axis $\alpha$ as defined in Eq.~\eqref{eq:sector_table} of dimension $16$. The charge tuples $\mathbf{I}=(I_1,\ldots,I_5)$ as defined in Eq.~\eqref{eq:allowed_charge_tuples} and the corresponding sector index $\alpha$ in the flat representation are listed in
Table~\ref{tab:z2_rank5_charge_tuples}.

\begin{table}[h]
  \centering
  \small
  \renewcommand{\arraystretch}{1.12}

  \begin{tabular*}{0.92\textwidth}{
    @{\extracolsep{\fill}}
    c c c c c c
    @{\hspace{0.8cm}}
    c c c c c c
  }
    \toprule
    $\alpha$
    & $I_1$ & $I_2$ & $I_3$ & $I_4$ & $I_5$
    &
    $\alpha$
    & $I_1$ & $I_2$ & $I_3$ & $I_4$ & $I_5$
    \\
    \midrule

     0 & 0 & 0 & 0 & 0 & 0
    &  8 & 1 & 0 & 0 & 0 & 1 \\

     1 & 0 & 0 & 0 & 1 & 1
    &  9 & 1 & 0 & 0 & 1 & 0 \\

     2 & 0 & 0 & 1 & 0 & 1
    & 10 & 1 & 0 & 1 & 0 & 0 \\

     3 & 0 & 0 & 1 & 1 & 0
    & 11 & 1 & 0 & 1 & 1 & 1 \\

     4 & 0 & 1 & 0 & 0 & 1
    & 12 & 1 & 1 & 0 & 0 & 0 \\

     5 & 0 & 1 & 0 & 1 & 0
    & 13 & 1 & 1 & 0 & 1 & 1 \\

     6 & 0 & 1 & 1 & 0 & 0
    & 14 & 1 & 1 & 1 & 0 & 1 \\

     7 & 0 & 1 & 1 & 1 & 1
    & 15 & 1 & 1 & 1 & 1 & 0 \\

    \bottomrule
  \end{tabular*}

  \caption{
    Charge tuples of the $16$ symmetry-allowed blocks of $\widetilde{X}$ and $\widetilde{Y}$.
    For $\widetilde{X}$, $(I_1,\ldots,I_5)=(I_a,I_b,I_c,I_u,I_v)$;
    for $\widetilde{Y}$, $(I_1,\ldots,I_5)=(I_{u},I_{v},I_d,I_e,I_f)$.
  }
  \label{tab:z2_rank5_charge_tuples}
\end{table}
\paragraph{Rearranging tensor blocks by sorting}
During contraction of block-sparse tensors, only the dense tensor blocks of
$\widetilde{X}$ and $\widetilde{Y}$ with matching charge tuples on the contracted legs $(u,v)$ are contracted to produce the output tensor blocks. 
In order to perform the block-sparse contraction in terms of dense tensor contraction, we align the blocks with matching charge tuples on the contracted legs, and rearrange the blocks in groups by the charge tuple of the outer (uncontracted) legs.

We first calculate the effective charge $q_C$ (as defined in Eq.~\eqref{eq:effective_charge}) on the contracted legs for each block from their charge tuple,
\begin{equation}
  q_C
  \equiv
  I_u+I_v
  \pmod 2,
  \qquad
  q_C\in\{0,1\}.
\end{equation}

We group the blocks by their $q_C$ and 
further factorize their charge tuples into outer-leg and contracted-leg charge tuples,
\begin{equation}
  \left.\mathcal{Q}_X\right|_{q_C}
  =
  \mathcal{L}_{q_C}\times\mathcal{C}_{q_C},
  \qquad
  \left.\mathcal{Q}_Y\right|_{q_C}
  =
  \mathcal{R}_{q_C}\times\mathcal{C}_{q_C}.
\end{equation}
as shown in Table~\ref{tab:z2_rank5_sector_sets}.
\begin{table}[h]
  \centering
  \small

  \renewcommand{\arraystretch}{1.18}

  \begin{tabular*}{0.72\textwidth}{
    @{\extracolsep{\fill}}
    c
    c
    c
  }
    \toprule
    $q_C$
    &
    $\mathcal{L}_{q_C}$ ($\mathcal{R}_{q_C}$)
    &
    $\mathcal{C}_{q_C}$
    \\
    \midrule

    $0$
    &
    $\{000,\;011,\;101,\;110\}$
    &
    $\{00,\;11\}$
    \\[2pt]

    $1$
    &
    $\{001,\;010,\;100,\;111\}$
    &
    $\{01,\;10\}$
    \\

    \bottomrule
  \end{tabular*}

  \caption{
    Decomposition of the charge tuples into outer-leg and contracted-leg charge tuples for different $q_C$ of the contracted legs.
  }
  \label{tab:z2_rank5_sector_sets}
\end{table}

We then arrange the blocks in lexicographic ordering according to the charge tuples
\begin{equation}
  \bigl(
    q_C,\mathbf{I}_L,\mathbf{I}_C
  \bigr),
  \qquad
  \bigl(
    q_C,\mathbf{I}_R,\mathbf{I}_{C}
  \bigr),
  \label{eq:XYtuples}
\end{equation}
where $\mathbf{I}_L\in\mathcal{L}$, $\mathbf{I}_R\in\mathcal{R}$ and $\mathbf{I}_C\in\mathcal{C}$ are charge tuples on the left, right outer legs and the contracted legs. For example as shown in Table.~\ref{tab:z2_rank5_sorted_X}, for $\widetilde{X}$, after the rearrangement, the blocks with the same $q_C$ are contiguous and in order ($q_C=0$ appears before $q_C=1$), then within each $q_C$ sector the blocks with the same $\mathbf{I}_L$ are contiguous, and finally within each $\mathbf{I}_L$ group the blocks are in order according to $\mathbf{I}_C$. The reordered sector axis has the
nested structure
\begin{equation}
  \underbrace{2}_{q_C}
  \times
  \underbrace{4}_{\mathbf{I}_L}
  \times
  \underbrace{2}_{\mathbf{I}_C}.
\end{equation}
For example, the first two consecutive blocks have
$(q_C,\mathbf{I}_L)=(0,000)$ and differ only in
$\mathbf{I}_C=00,11$, while the next two have
$(q_C,\mathbf{I}_L)=(0,011)$ and again differ only in
$\mathbf{I}_C=00,11$.
Therefore the reordered linear sector index $\tilde{\alpha}$ can be
reshaped directly into the multi-index $(q_C,\lambda,\gamma)$ of shape
$2\times4\times2$.

\begin{table}[h]
  \centering
  \small
  \renewcommand{\arraystretch}{1.18}

  \begin{tabular*}{0.82\textwidth}{
    @{\extracolsep{\fill}}
    c c c c c
  }
    \toprule
    sorted index $\tilde{\alpha}$
    & $q_C$
    & $\mathbf{I}_L$
    & $\mathbf{I}_C$
    & original $\alpha$
    \\
    \midrule

     0 & 0 & $000$ & $00$ & 0  \\
     1 & 0 & $000$ & $11$ & 1  \\
     2 & 0 & $011$ & $00$ & 6  \\
     3 & 0 & $011$ & $11$ & 7  \\
     4 & 0 & $101$ & $00$ & 10 \\
     5 & 0 & $101$ & $11$ & 11 \\
     6 & 0 & $110$ & $00$ & 12 \\
     7 & 0 & $110$ & $11$ & 13 \\

    \midrule

     8 & 1 & $001$ & $01$ & 2  \\
     9 & 1 & $001$ & $10$ & 3  \\
    10 & 1 & $010$ & $01$ & 4  \\
    11 & 1 & $010$ & $10$ & 5  \\
    12 & 1 & $100$ & $01$ & 8  \\
    13 & 1 & $100$ & $10$ & 9  \\
    14 & 1 & $111$ & $01$ & 14 \\
    15 & 1 & $111$ & $10$ & 15 \\

    \bottomrule
  \end{tabular*}

  \caption{
    Charge sectors of $\widetilde{X}$ arranged lexicographically according to
    the composite key $(q_C,\mathbf{I}_L,\mathbf{I}_C)$.
    The new sector index $\tilde{\alpha}$ labels positions along the reordered
    sector axis, while $\alpha$ denotes the sector index in the original
    ordering of Table~\ref{tab:z2_rank5_charge_tuples}.
  }
  \label{tab:z2_rank5_sorted_X}
\end{table}

For $\widetilde{X}$ and $\widetilde{Y}$, we have
\begin{equation}
  \widetilde{\alpha}_X
  \longleftrightarrow
  (q_C,\lambda,\gamma^{X}),
  \qquad
  \widetilde{\alpha}_Y
  \longleftrightarrow
  (q_C,\rho,\gamma^{Y}),
  \label{eq:z2_rank5_sector_reindex}
\end{equation}
where
\begin{equation}
  q_C\in \{0,1\},
  \qquad
  \lambda,\rho\in\{0,\ldots,3\},
  \qquad
  \gamma^{X},\gamma^{Y}\in\{0,1\}.
\end{equation}
Here $\lambda$ enumerates the four tuples in $\mathcal{L}_{q_C}$,
$\rho$ enumerates the four tuples in $\mathcal{R}_{q_C}$, and
$\gamma^X$ and $\gamma^Y$ enumerate the two tuples in $\mathcal{C}_{q_C}$.

For $\widetilde{X}$, the blocks with $q_C=0$ are arranged as
a $4\times2$ matrix:
\begin{equation}
  \widetilde{X}^{(q_C=0)}
  =
  \begin{pmatrix}
    \widetilde{X}_{(000,00)} & \widetilde{X}_{(000,11)} \\
    \widetilde{X}_{(011,00)} & \widetilde{X}_{(011,11)} \\
    \widetilde{X}_{(101,00)} & \widetilde{X}_{(101,11)} \\
    \widetilde{X}_{(110,00)} & \widetilde{X}_{(110,11)}
  \end{pmatrix},
  \label{eq:z2_rank5_X_block_table}
\end{equation}
where each element corresponds to the dense tensor block $\widetilde{X}_{(\mathbf{I}_L,{\mathbf{I}_C})}$, and the row and column index correspond to $\lambda$ and $\gamma^X$, respectively.
Likewise, the blocks of $\widetilde{Y}$ after transposition, displayed with the $\gamma^Y$ 
as the row index and $\rho$ as the column index, are arranged as a $2\times 4$ matrix:
\begin{equation}
  \left(\widetilde{Y}^{(q_C=0)}\right)^{\mathsf T}
  =
  \begin{pmatrix}
    \widetilde{Y}_{(00,000)}
    &
    \widetilde{Y}_{(00,011)}
    &
    \widetilde{Y}_{(00,101)}
    &
    \widetilde{Y}_{(00,110)}
    \\
    \widetilde{Y}_{(11,000)}
    &
    \widetilde{Y}_{(11,011)}
    &
    \widetilde{Y}_{(11,101)}
    &
    \widetilde{Y}_{(11,110)}
  \end{pmatrix}.
  \label{eq:z2_rank5_Y_block_table}
\end{equation}

\paragraph{Block-sparse tensor contraction as dense tensor contraction in the flat representation.}
Since the tensor blocks are arranged according to
$(q_C,\mathbf{I}_L,\mathbf{I}_C)$ before reshaping the leading sector axis
into $(q_C,\lambda,\gamma)$, a target output block
$\widetilde{Z}_{q_C,\lambda,\rho}$ and its contributing input blocks
$\widetilde{X}_{q_C,\lambda,\gamma}$ and
$\widetilde{Y}_{q_C,\rho,\gamma}$ can be selected directly through slicing the
indices $(q_C,\lambda,\rho)$, with the contraction performed by summing over
$\gamma$, as illustrated in
Fig.~\ref{fig:sector-sort-reshape-contract}(e). Suppressing the dense intra-sector indices, the contraction to obtain blocks in $\widetilde{Z}$ is written as:
\begin{equation}
  \widetilde{Z}^{(q_C)}_{\lambda\rho}
  =
  \sum_{\gamma=0}^{1}
  C\Bigl(\widetilde{X}^{(q_C)}_{\lambda\gamma},
  (\widetilde{Y}^{(q_C)}_{\rho\gamma})^\mathsf T\Bigr).
  \label{eq:z2_rank5_channel_contraction}
\end{equation}
where $C(\cdot,\cdot)$ denotes contraction over the intra-sector indices of the dense tensor block. Each tuple of $(q_C,\lambda,\rho)$ uniquely maps to a valid tensor block of $\widetilde{Z}$ with charge tuple $(\mathbf{I}_L,\mathbf{I}_R)$. For example, $(q_C=0,\lambda=1,\rho=2)$ corresponds to the output tensor block with charge tuple $(\mathbf{I}_L,\mathbf{I}_R)=(011,101)$ and is obtained by the following dense tensor contraction:
\begin{equation}
  \begin{aligned}
  \widetilde{Z}_{(011,101)}
  =
    C\Bigl(\widetilde{X}_{(011,{\color{red}00})},
    \widetilde{Y}^{\mathsf T}_{({\color{red}00},101)}\Bigr)
    +
    C\Bigl(\widetilde{X}_{(011,{\color{red}11})},
    \widetilde{Y}^{\mathsf T}_{({\color{red}11},101)}\Bigr).
  \end{aligned}
  \label{eq:z2_rank5_example_block}
\end{equation}

For each value of $q_C$, the contraction is equivalent to multiplying
the $4\times2$ dense tensor block matrix Eq.~\eqref{eq:z2_rank5_X_block_table} by the $2\times4$ dense tensor block matrix Eq.~\eqref{eq:z2_rank5_Y_block_table}, giving a total of $4\times4=16$ output tensor blocks.
Combining the blocks for both $q_C=0$ and $q_C=1$, we obtain $
  2\times4\times4
  =
  32
  =
  2^{6-1}
$
output blocks, as required for a sector-complete rank-six
$\mathbb{Z}_2$ tensor with total charge zero.

Therefore, the contraction of two block-sparse tensors $\widetilde{X}$ and $\widetilde{Y}$ to obtain the output tensor $\widetilde{Z}$ can be directly performed as the dense tensor contraction of $\widetilde{X}$ and $\widetilde{Y}$ in the flat representation, and the output tensor $\widetilde{Z}$ is also automatically in the flat representation suitable for subsequent block-sparse tensor operations represented as dense operations.

\section{Vectorizing standard fPEPS-VMC with boundary-MPS environment reuse}
\label{app:bmps_reuse}

\begin{figure}[!htbp]
\centering
\begin{subfigure}[!htbp]{0.48\textwidth}
  \centering
  \includegraphics[width=0.62\columnwidth]{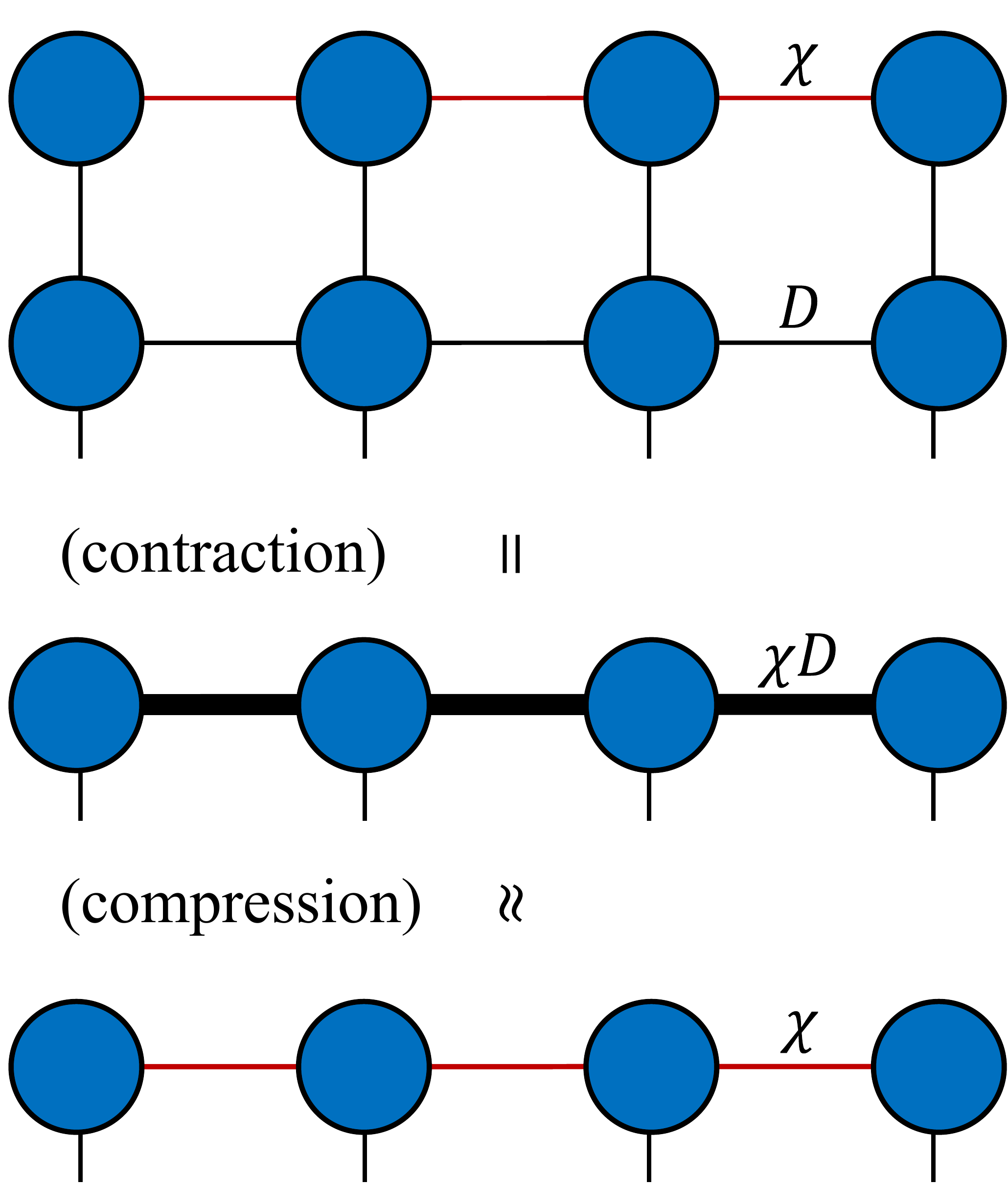}
  \caption{Boundary-MPS contraction.}
  \label{fig:bmps_contraction}
\end{subfigure}
\hfill
\begin{subfigure}[!htbp]{0.48\textwidth}
  \centering
  \includegraphics[width=0.85\columnwidth]{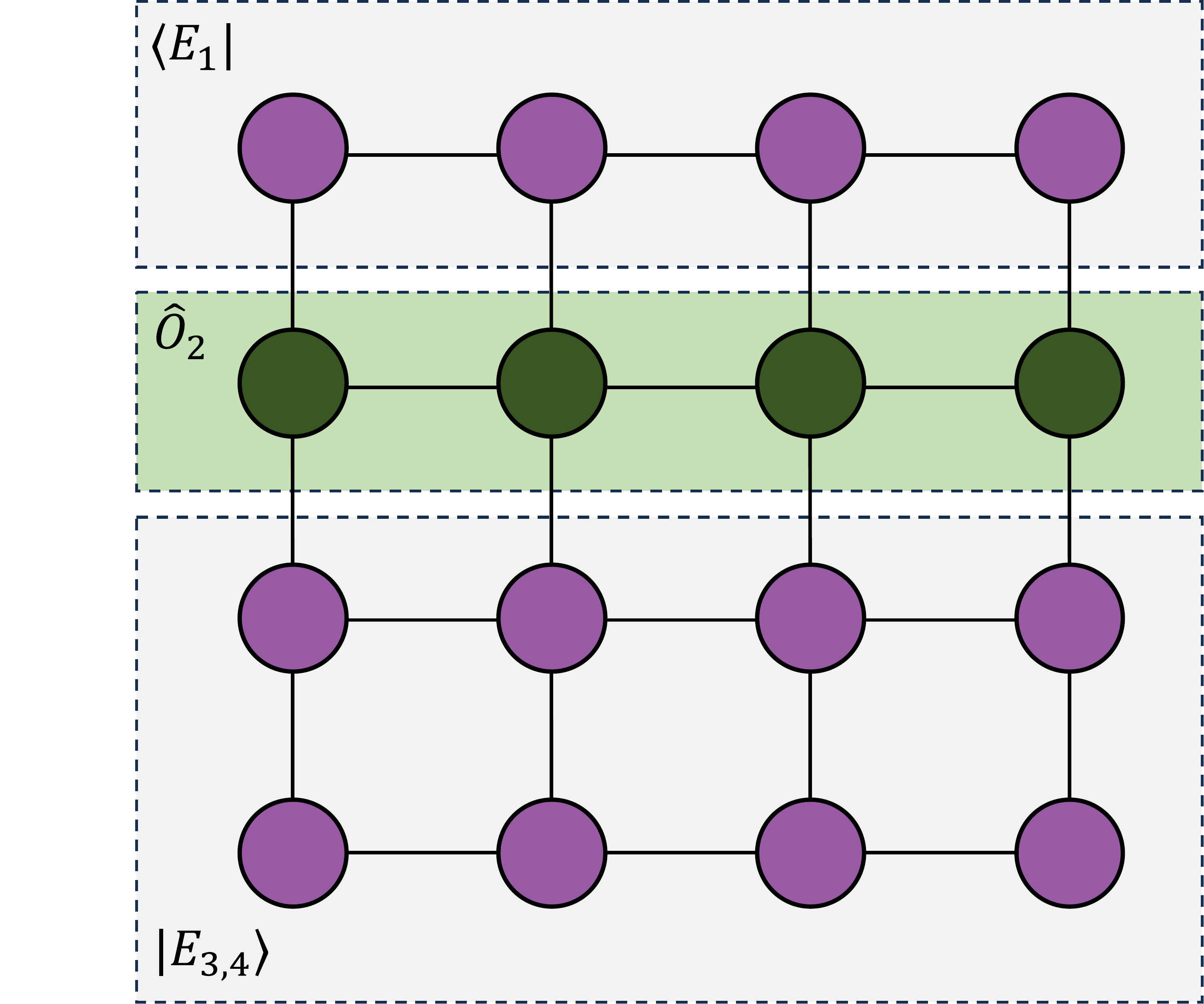}
  \caption{Boundary-MPS environment reuse.}
  \label{fig:bmps_reuse}
\end{subfigure}
\caption{\textbf{Boundary-MPS contraction and environment reuse for standard
fPEPS amplitude evaluation.} \textbf{(a)} Boundary-MPS contraction proceeds row
by row. Absorbing a PEPS row with virtual bond dimension $D$ into a boundary
MPS with bond dimension $\chi$ temporarily increases the boundary-MPS bond
dimension to $\chi D$; the boundary MPS is then compressed back to bond
dimension $\chi$. \textbf{(b)} For amplitudes that differ from a reference
configuration only within a local region, the boundary environments outside the
active region can be reused. In the example shown, the amplitude is written as
$\Psi=\langle E_1|\hat{O}_2|E_{3,4}\rangle$, where $\langle E_1|$ and
$|E_{3,4}\rangle$ are cached environments and $\hat{O}_2$ is the active row.
Configurations with the same active-row geometry share the same residual
contraction computational graph and can therefore be evaluated in a single vectorized batch,
with the environments and active-row tensors supplied as batched data.}
\label{fig:bmps_reuse_appendix}
\end{figure}

We describe how boundary-MPS environment reuse in standard fPEPS-VMC can be
combined with vectorized amplitude evaluation. A projected PEPS amplitude is
computed by absorbing the PEPS rows of bond dimension $D$ sequentially into a boundary MPS of bond dimension $\chi$. After
each row is absorbed, the boundary-MPS with increased bond dimension $D\chi$ is compressed back to a smaller MPS of bond dimension
$\chi$, as illustrated in Fig.~\ref{fig:bmps_contraction}.

In a VMC calculation, many amplitudes correspond to configurations that differ
from a reference configuration only through local changes. These include trial
configurations generated during Metropolis sampling and configurations connected
to a sampled configuration by local Hamiltonian terms in the local-energy
evaluation. Because the (fermionic) PEPS is composed of local site tensors, changing the
physical configuration modifies only the corresponding projected tensors in
the amplitude network. The remaining parts of the network can therefore be
represented by cached boundary environments.

For example, when the modified sites lie in row~2, the rows above and below can
be contracted into the cached environments $\langle E_1|$ and
$|E_{3,4}\rangle$, respectively. The amplitude is then obtained from contracting the residual tensor network
\begin{equation}
  \Psi
  =
  \langle E_1|\hat{O}_2|E_{3,4}\rangle ,
  \label{eq:bmps_reuse_row}
\end{equation}
where $\hat{O}_2$ denotes the active row
(Fig.~\ref{fig:bmps_reuse}). Combined with a deterministic sequential update
scheme in the sampler, this environment-reuse strategy reduces the cost per
VMC step from $O(N^2)$ to $O(N)$ for standard
PEPS-VMC~\cite{Liu_2017,Liu2021accuratepeps}, where
$N=L_xL_y$ is the number of lattice sites.

For vectorization over the physical configurations $\sigma$, the relevant observation is that the residual contraction
graph is determined by the position of the modified sites rather than by the
configuration values on the modified sites. Configurations whose changes are
confined to the same active row therefore produce residual tensor networks with the same geometry. Their amplitudes can be evaluated
as a single batch by stacking the cached top and bottom environments and the
active-row tensors along a leading batch dimension and applying the shared
contraction computational graph.

Unlike in the TNF-fPEPS case, boundary-MPS reuse therefore does not produce one global common contraction computational graph for
all amplitude evaluations.  
Instead, it yields a
finite collection of computational graphs indexed by the geometry of the residual tensor network. 
With row-environment reuse, the graph is indexed by the modified row,
resulting in $O(L_x)$ possible geometries. If intra-row left and right
environments are also reused, the residual network can be localized around a few consecutive sites, resulting in
$O(L_xL_y)=O(N)$ possible geometries.

Within each geometry class, however, all amplitude evaluations share the same
computational graph and can be vectorized as described in the main text. The
standard fPEPS-VMC workflow with boundary-MPS reuse is therefore implemented by
grouping amplitude evaluations according to their residual network geometry and
evaluating each group as a vectorized tensor-network contraction.

\end{appendix}

\bibliography{references}

@article{mcmillan1965ground,
  title = {Ground State of Liquid ${\mathrm{He}}^{4}$},
  author = {McMillan, W. L.},
  journal = {Phys. Rev.},
  volume = {138},
  issue = {2A},
  pages = {A442--A451},
  numpages = {0},
  year = {1965},
  month = {Apr},
  publisher = {American Physical Society},
  doi = {10.1103/PhysRev.138.A442},
  url = {https://link.aps.org/doi/10.1103/PhysRev.138.A442}
}

@article{ceperley1977monte,
  title = {Monte Carlo simulation of a many-fermion study},
  author = {Ceperley, D. and Chester, G. V. and Kalos, M. H.},
  journal = {Phys. Rev. B},
  volume = {16},
  issue = {7},
  pages = {3081--3099},
  numpages = {0},
  year = {1977},
  month = {Oct},
  publisher = {American Physical Society},
  doi = {10.1103/PhysRevB.16.3081},
  url = {https://link.aps.org/doi/10.1103/PhysRevB.16.3081}
}

@article{foulkes2001quantum,
  title = {Quantum Monte Carlo simulations of solids},
  author = {Foulkes, W. M. C. and Mitas, L. and Needs, R. J. and Rajagopal, G.},
  journal = {Rev. Mod. Phys.},
  volume = {73},
  issue = {1},
  pages = {33--83},
  numpages = {0},
  year = {2001},
  month = {Jan},
  publisher = {American Physical Society},
  doi = {10.1103/RevModPhys.73.33},
  url = {https://link.aps.org/doi/10.1103/RevModPhys.73.33}
}

@book{Ran_2020,
   title={Tensor Network Contractions: Methods and Applications to Quantum Many-Body Systems},
   ISBN={9783030344894},
   ISSN={1616-6361},
   url={http://dx.doi.org/10.1007/978-3-030-34489-4},
   DOI={10.1007/978-3-030-34489-4},
   journal={Lecture Notes in Physics},
   publisher={Springer International Publishing},
   author={Ran, Shi-Ju and Tirrito, Emanuele and Peng, Cheng and Chen, Xi and Tagliacozzo, Luca and Su, Gang and Lewenstein, Maciej},
   year={2020} }

@article{white1992density,
  title = {Density matrix formulation for quantum renormalization groups},
  author = {White, Steven R.},
  journal = {Phys. Rev. Lett.},
  volume = {69},
  issue = {19},
  pages = {2863--2866},
  numpages = {0},
  year = {1992},
  month = {Nov},
  publisher = {American Physical Society},
  doi = {10.1103/PhysRevLett.69.2863},
  url = {https://link.aps.org/doi/10.1103/PhysRevLett.69.2863}
}

@misc{chen2025tensornetworkmarkovchain,
      title={Tensor Network Markov Chain Monte Carlo: Efficient Sampling of Three-Dimensional Spin Glasses and Beyond}, 
      author={Tao Chen and Jing Liu and Youjin Deng and Pan Zhang},
      year={2025},
      eprint={2509.23945},
      archivePrefix={arXiv},
      primaryClass={cond-mat.stat-mech},
      url={https://arxiv.org/abs/2509.23945}, 
}

@misc{park2026statisticalmechanicscontinuousspace,
      title={Statistical mechanics in continuous space with tensor network methods}, 
      author={Gunhee Park and Tomislav Begušić and Si-Jing Du and Johnnie Gray and Garnet Kin-Lic Chan},
      year={2026},
      eprint={2604.25060},
      archivePrefix={arXiv},
      primaryClass={cond-mat.stat-mech},
      url={https://arxiv.org/abs/2604.25060}, 
}

@misc{chen2025batchtnmcefficientsamplingtwodimensional,
      title={BatchTNMC: Efficient sampling of two-dimensional spin glasses using tensor network Monte Carlo}, 
      author={Tao Chen and Jingtong Zhang and Jing Liu and Youjin Deng and Pan Zhang},
      year={2025},
      eprint={2509.19006},
      archivePrefix={arXiv},
      primaryClass={cond-mat.stat-mech},
      url={https://arxiv.org/abs/2509.19006}, 
}

@article{Liu2021accuratepeps,
  title = {Accurate simulation for finite projected entangled pair states in two dimensions},
  author = {Liu, Wen-Yuan and Huang, Yi-Zhen and Gong, Shou-Shu and Gu, Zheng-Cheng},
  journal = {Phys. Rev. B},
  volume = {103},
  issue = {23},
  pages = {235155},
  numpages = {13},
  year = {2021},
  month = {Jun},
  publisher = {American Physical Society},
  doi = {10.1103/PhysRevB.103.235155},
  url = {https://link.aps.org/doi/10.1103/PhysRevB.103.235155}
}

@article{Kraus_2010,
   title={Fermionic projected entangled pair states},
   volume={81},
   ISSN={1094-1622},
   url={http://dx.doi.org/10.1103/PhysRevA.81.052338},
   DOI={10.1103/physreva.81.052338},
   number={5},
   journal={Physical Review A},
   publisher={American Physical Society (APS)},
   author={Kraus, Christina V. and Schuch, Norbert and Verstraete, Frank and Cirac, J. Ignacio},
   year={2010},
   month=May }

@article{Pi_orn_2010,
   title={Fermionic implementation of projected entangled pair states algorithm},
   volume={81},
   ISSN={1550-235X},
   url={http://dx.doi.org/10.1103/PhysRevB.81.245110},
   DOI={10.1103/physrevb.81.245110},
   number={24},
   journal={Physical Review B},
   publisher={American Physical Society (APS)},
   author={Pižorn, Iztok and Verstraete, Frank},
   year={2010},
   month=June }

@article{Singh_2010,
   title={Tensor network decompositions in the presence of a global symmetry},
   volume={82},
   ISSN={1094-1622},
   url={http://dx.doi.org/10.1103/PhysRevA.82.050301},
   DOI={10.1103/physreva.82.050301},
   number={5},
   journal={Physical Review A},
   publisher={American Physical Society (APS)},
   author={Singh, Sukhwinder and Pfeifer, Robert N. C. and Vidal, Guifré},
   year={2010},
   month=Nov }

@article{Singh_2011,
   title={Tensor network states and algorithms in the presence of a global U(1) symmetry},
   volume={83},
   ISSN={1550-235X},
   url={http://dx.doi.org/10.1103/PhysRevB.83.115125},
   DOI={10.1103/physrevb.83.115125},
   number={11},
   journal={Physical Review B},
   publisher={American Physical Society (APS)},
   author={Singh, Sukhwinder and Pfeifer, Robert N. C. and Vidal, Guifre},
   year={2011},
   month=Mar }

@article{Bauer_2011,
   title={Implementing global Abelian symmetries in projected entangled-pair state algorithms},
   volume={83},
   ISSN={1550-235X},
   url={http://dx.doi.org/10.1103/PhysRevB.83.125106},
   DOI={10.1103/physrevb.83.125106},
   number={12},
   journal={Physical Review B},
   publisher={American Physical Society (APS)},
   author={Bauer, B. and Corboz, P. and Orús, R. and Troyer, M.},
   year={2011},
   month=Mar }

@article{Corboz_2010,
   title={Simulation of strongly correlated fermions in two spatial dimensions with fermionic projected entangled-pair states},
   volume={81},
   ISSN={1550-235X},
   url={http://dx.doi.org/10.1103/PhysRevB.81.165104},
   DOI={10.1103/physrevb.81.165104},
   number={16},
   journal={Physical Review B},
   publisher={American Physical Society (APS)},
   author={Corboz, Philippe and Orús, Román and Bauer, Bela and Vidal, Guifré},
   year={2010},
   month=Apr }

@article{Cirac_2021,
   title={Matrix product states and projected entangled pair states: Concepts, symmetries, theorems},
   volume={93},
   ISSN={1539-0756},
   url={http://dx.doi.org/10.1103/RevModPhys.93.045003},
   DOI={10.1103/revmodphys.93.045003},
   number={4},
   journal={Reviews of Modern Physics},
   publisher={American Physical Society (APS)},
   author={Cirac, J. Ignacio and Pérez-García, David and Schuch, Norbert and Verstraete, Frank},
   year={2021},
   month=Dec }

@misc{gu2010grassmanntensornetworkstates,
      title={Grassmann tensor network states and its renormalization for strongly correlated fermionic and bosonic states}, 
      author={Zheng-Cheng Gu and Frank Verstraete and Xiao-Gang Wen},
      year={2010},
      eprint={1004.2563},
      archivePrefix={arXiv},
      primaryClass={cond-mat.str-el},
      url={https://arxiv.org/abs/1004.2563}, 
}

@article{schollwock2011density,
   title={The density-matrix renormalization group in the age of matrix product states},
   volume={326},
   ISSN={0003-4916},
   url={http://dx.doi.org/10.1016/j.aop.2010.09.012},
   DOI={10.1016/j.aop.2010.09.012},
   number={1},
   journal={Annals of Physics},
   publisher={Elsevier BV},
   author={Schollwöck, Ulrich},
   year={2011},
   month=Jan, pages={96–192} }

@misc{verstraete2004renormalization,
      title={Renormalization algorithms for Quantum-Many Body Systems in two and higher dimensions}, 
      author={F. Verstraete and J. I. Cirac},
      year={2004},
      eprint={cond-mat/0407066},
      archivePrefix={arXiv},
      primaryClass={cond-mat.str-el},
      url={https://arxiv.org/abs/cond-mat/0407066}, 
}

@article{corboz2016variational,
   title={Variational optimization with infinite projected entangled-pair states},
   volume={94},
   ISSN={2469-9969},
   url={http://dx.doi.org/10.1103/PhysRevB.94.035133},
   DOI={10.1103/physrevb.94.035133},
   number={3},
   journal={Physical Review B},
   publisher={American Physical Society (APS)},
   author={Corboz, Philippe},
   year={2016},
   month=July }

@article{du2024neuralized,
  title = {Neuralized fermionic tensor networks for quantum many-body systems},
  author = {Du, Si-Jing and Chen, Ao and Chan, Garnet Kin-Lic},
  journal = {Phys. Rev. B},
  volume = {113},
  issue = {8},
  pages = {085134},
  numpages = {13},
  year = {2026},
  month = {Feb},
  publisher = {American Physical Society},
  doi = {10.1103/x8vl-qf14},
  url = {https://link.aps.org/doi/10.1103/x8vl-qf14}
}

@article{Robledo_Moreno_2022,
   title={Fermionic wave functions from neural-network constrained hidden states},
   volume={119},
   ISSN={1091-6490},
   url={http://dx.doi.org/10.1073/pnas.2122059119},
   DOI={10.1073/pnas.2122059119},
   number={32},
   journal={Proceedings of the National Academy of Sciences},
   publisher={National Academy of Sciences},
   author={Robledo Moreno, Javier and Carleo, Giuseppe and Georges, Antoine and Stokes, James},
   year={2022},
   month=Aug }

@article{Or_s_2014,
   title={A practical introduction to tensor networks: Matrix product states and projected entangled pair states},
   volume={349},
   ISSN={0003-4916},
   url={http://dx.doi.org/10.1016/j.aop.2014.06.013},
   DOI={10.1016/j.aop.2014.06.013},
   journal={Annals of Physics},
   publisher={Elsevier BV},
   author={Orús, Román},
   year={2014},
   month=Oct, pages={117–158} }

@article{Metropolis_1953,
    author = {Metropolis, Nicholas and Rosenbluth, Arianna W. and Rosenbluth, Marshall N. and Teller, Augusta H. and Teller, Edward},
    title = {Equation of State Calculations by Fast Computing Machines},
    journal = {The Journal of Chemical Physics},
    volume = {21},
    number = {6},
    pages = {1087-1092},
    year = {1953},
    month = {06},
    issn = {0021-9606},
    doi = {10.1063/1.1699114},
    url = {https://doi.org/10.1063/1.1699114},
    eprint = {https://pubs.aip.org/aip/jcp/article-pdf/21/6/1087/18802390/1087_1_online.pdf},
}

@article{Hastings_1970,
 ISSN = {00063444, 14643510},
 URL = {http://www.jstor.org/stable/2334940},
 author = {W. K. Hastings},
 journal = {Biometrika},
 number = {1},
 pages = {97--109},
 publisher = {[Oxford University Press, Biometrika Trust]},
 title = {Monte Carlo Sampling Methods Using Markov Chains and Their Applications},
 urldate = {2026-06-29},
 volume = {57},
 year = {1970}
}

@article{Liu_2017,
   title={Gradient optimization of finite projected entangled pair states},
   volume={95},
   ISSN={2469-9969},
   url={http://dx.doi.org/10.1103/PhysRevB.95.195154},
   DOI={10.1103/physrevb.95.195154},
   number={19},
   journal={Physical Review B},
   publisher={American Physical Society (APS)},
   author={Liu, Wen-Yuan and Dong, Shao-Jun and Han, Yong-Jian and Guo, Guang-Can and He, Lixin},
   year={2017},
   month=May }

@article{Schmitt_2022,
   title={jVMC: Versatile and performant variational Monte Carlo leveraging  automated differentiation and GPU acceleration},
   url={http://dx.doi.org/10.21468/SciPostPhysCodeb.2},
   DOI={10.21468/scipostphyscodeb.2},
   journal={SciPost Physics Codebases},
   publisher={Stichting SciPost},
   author={Schmitt, Markus and Reh, Moritz},
   year={2022},
   month=Aug }

@article{Begu_i__2024,
   title={Fast and converged classical simulations of evidence for the utility of quantum computing before fault tolerance},
   volume={10},
   ISSN={2375-2548},
   url={http://dx.doi.org/10.1126/sciadv.adk4321},
   DOI={10.1126/sciadv.adk4321},
   number={3},
   journal={Science Advances},
   publisher={American Association for the Advancement of Science (AAAS)},
   author={Begušić, Tomislav and Gray, Johnnie and Chan, Garnet Kin-Lic},
   year={2024},
   month=Jan }

@misc{pan2024efficientquantumcircuitsimulation,
      title={Efficient Quantum Circuit Simulation by Tensor Network Methods on Modern GPUs}, 
      author={Feng Pan and Hanfeng Gu and Lvlin Kuang and Bing Liu and Pan Zhang},
      year={2024},
      eprint={2310.03978},
      archivePrefix={arXiv},
      primaryClass={quant-ph},
      url={https://arxiv.org/abs/2310.03978}, 
}

@article{pancircuit2022,
  title = {Simulation of Quantum Circuits Using the Big-Batch Tensor Network Method},
  author = {Pan, Feng and Zhang, Pan},
  journal = {Phys. Rev. Lett.},
  volume = {128},
  issue = {3},
  pages = {030501},
  numpages = {6},
  year = {2022},
  month = {Jan},
  publisher = {American Physical Society},
  doi = {10.1103/PhysRevLett.128.030501},
  url = {https://link.aps.org/doi/10.1103/PhysRevLett.128.030501}
}

@article{Pan_2022,
   title={Solving the Sampling Problem of the Sycamore Quantum Circuits},
   volume={129},
   ISSN={1079-7114},
   url={http://dx.doi.org/10.1103/PhysRevLett.129.090502},
   DOI={10.1103/physrevlett.129.090502},
   number={9},
   journal={Physical Review Letters},
   publisher={American Physical Society (APS)},
   author={Pan, Feng and Chen, Keyang and Zhang, Pan},
   year={2022},
   month=Aug }

@article{Liu_2021_tropicaltensor,
   title={Tropical Tensor Network for Ground States of Spin Glasses},
   volume={126},
   ISSN={1079-7114},
   url={http://dx.doi.org/10.1103/PhysRevLett.126.090506},
   DOI={10.1103/physrevlett.126.090506},
   number={9},
   journal={Physical Review Letters},
   publisher={American Physical Society (APS)},
   author={Liu, Jin-Guo and Wang, Lei and Zhang, Pan},
   year={2021},
   month=Mar }

@article{Liu_2023_combinatorialoptimization,
   title={Computing Solution Space Properties of Combinatorial Optimization Problems Via Generic Tensor Networks},
   volume={45},
   ISSN={1095-7197},
   url={http://dx.doi.org/10.1137/22M1501787},
   DOI={10.1137/22m1501787},
   number={3},
   journal={SIAM Journal on Scientific Computing},
   publisher={Society for Industrial & Applied Mathematics (SIAM)},
   author={Liu, Jin-Guo and Gao, Xun and Cain, Madelyn and Lukin, Mikhail D. and Wang, Sheng-Tao},
   year={2023},
   month=June, pages={A1239–A1270} }

@article{Pan_2020,
   title={Contracting Arbitrary Tensor Networks: General Approximate Algorithm and Applications in Graphical Models and Quantum Circuit Simulations},
   volume={125},
   ISSN={1079-7114},
   url={http://dx.doi.org/10.1103/PhysRevLett.125.060503},
   DOI={10.1103/physrevlett.125.060503},
   number={6},
   journal={Physical Review Letters},
   publisher={American Physical Society (APS)},
   author={Pan, Feng and Zhou, Pengfei and Li, Sujie and Zhang, Pan},
   year={2020},
   month=Aug }

@article{Huang_2021nsy,
    author = "Huang, Cupjin and others",
    title = "{Efficient parallelization of tensor network contraction for simulating quantum computation}",
    doi = "10.1038/s43588-021-00119-7",
    journal = "Nature Computat. Sci.",
    volume = "1",
    pages = "578--587",
    year = "2021"
}

@software{jax2018github,
  author = {James Bradbury and Roy Frostig and Peter Hawkins and Matthew James Johnson and Yash Katariya and Chris Leary and Dougal Maclaurin and George Necula and Adam Paszke and Jake Vander{P}las and Skye Wanderman-{M}ilne and Qiao Zhang},
  title = {{JAX}: composable transformations of {P}ython+{N}um{P}y programs},
  url = {http://github.com/jax-ml/jax},
  version = {0.3.13},
  year = {2018},
}

@article{torch2019github,
  author       = {Adam Paszke and
                  Sam Gross and
                  Francisco Massa and
                  Adam Lerer and
                  James Bradbury and
                  Gregory Chanan and
                  Trevor Killeen and
                  Zeming Lin and
                  Natalia Gimelshein and
                  Luca Antiga and
                  Alban Desmaison and
                  Andreas K{\"{o}}pf and
                  Edward Z. Yang and
                  Zach DeVito and
                  Martin Raison and
                  Alykhan Tejani and
                  Sasank Chilamkurthy and
                  Benoit Steiner and
                  Lu Fang and
                  Junjie Bai and
                  Soumith Chintala},
  title        = {PyTorch: An Imperative Style, High-Performance Deep Learning Library},
  journal      = {CoRR},
  volume       = {abs/1912.01703},
  year         = {2019},
  url          = {http://arxiv.org/abs/1912.01703},
  eprinttype   = {arXiv},
  eprint       = {1912.01703},
  bibsource    = {dblp computer science bibliography, https://dblp.org}
}

@misc{gu2025solvinghubbardmodelneural,
      title={Solving the Hubbard model with Neural Quantum States}, 
      author={Yuntian Gu and Wenrui Li and Heng Lin and Bo Zhan and Ruichen Li and Yifei Huang and Di He and Yantao Wu and Tao Xiang and Mingpu Qin and Liwei Wang and Dingshun Lv},
      year={2025},
      eprint={2507.02644},
      archivePrefix={arXiv},
      primaryClass={cond-mat.str-el},
      url={https://arxiv.org/abs/2507.02644}, 
}

@misc{fan2026nntns,
      title={Disentangling Tensor Network States with Deep Neural Network}, 
      author={Chaohui Fan and Bo Zhan and Yuntian Gu and Tong Liu and Yantao Wu and Mingpu Qin and Dingshun Lv and Tao Xiang},
      year={2026},
      eprint={2603.14425},
      archivePrefix={arXiv},
      primaryClass={cond-mat.str-el},
      url={https://arxiv.org/abs/2603.14425}, 
}

@article{Gray2021hyperoptimized,
  doi = {10.22331/q-2021-03-15-410},
  url = {https://doi.org/10.22331/q-2021-03-15-410},
  title = {Hyper-optimized tensor network contraction},
  author = {Gray, Johnnie and Kourtis, Stefanos},
  journal = {{Quantum}},
  issn = {2521-327X},
  publisher = {{Verein zur F{\"{o}}rderung des Open Access Publizierens in den Quantenwissenschaften}},
  volume = {5},
  pages = {410},
  month = mar,
  year = {2021}
}

@article{Filippo2022netket,
	title={{NetKet 3: Machine Learning Toolbox for Many-Body Quantum Systems}},
	author={Filippo Vicentini and Damian Hofmann and Attila Szabó and Dian Wu and Christopher   Roth and Clemens Giuliani and Gabriel Pescia and Jannes Nys and Vladimir Vargas-Calderón and   Nikita Astrakhantsev and Giuseppe Carleo},
	journal={SciPost Phys. Codebases},
	pages={7},
	year={2022},
	publisher={SciPost},
	doi={10.21468/SciPostPhysCodeb.7},
	url={https://scipost.org/10.21468/SciPostPhysCodeb.7},
}

@article{Gray2024hyperoptimized,
  title = {Hyperoptimized Approximate Contraction of Tensor Networks with Arbitrary Geometry},
  author = {Gray, Johnnie and Chan, Garnet Kin-Lic},
  journal = {Phys. Rev. X},
  volume = {14},
  issue = {1},
  pages = {011009},
  numpages = {19},
  year = {2024},
  month = {Jan},
  publisher = {American Physical Society},
  doi = {10.1103/PhysRevX.14.011009},
  url = {https://link.aps.org/doi/10.1103/PhysRevX.14.011009}
}

@article{liu2024tensor,
   title={Tensor Network Computations That Capture Strict Variationality, Volume Law Behavior, and the Efficient Representation of Neural Network States},
   volume={133},
   ISSN={1079-7114},
   url={http://dx.doi.org/10.1103/PhysRevLett.133.260404},
   DOI={10.1103/physrevlett.133.260404},
   number={26},
   journal={Physical Review Letters},
   publisher={American Physical Society (APS)},
   author={Liu, Wen-Yuan and Du, Si-Jing and Peng, Ruojing and Gray, Johnnie and Chan, Garnet Kin-Lic},
   year={2024},
   month=Dec }

@article{gao2024fermionic,
  title = {Fermionic tensor network contraction for arbitrary geometries},
  author = {Gao, Yang and Zhai, Huanchen and Gray, Johnnie and Peng, Ruojing and Park, Gunhee and Liu, Wen-Yuan and Kj\o{}nstad, Eirik F. and Chan, Garnet Kin-Lic},
  journal = {Phys. Rev. Res.},
  volume = {7},
  issue = {2},
  pages = {023193},
  numpages = {10},
  year = {2025},
  month = {May},
  publisher = {American Physical Society},
  doi = {10.1103/PhysRevResearch.7.023193},
  url = {https://link.aps.org/doi/10.1103/PhysRevResearch.7.023193}
}

@article{SciPostPhysCodeb.10,
	title = {Automatic transformation of irreducible representations for efficient contraction of tensors with cyclic group symmetry},
	pages = {10},
	author = {Gao, Yang and Helms, Phillip and Chan, Garnet Kin-Lic and Solomonik, Edgar},
	journal = {SciPost Phys. Codebases},
	year = {2023},
	publisher = {SciPost},
	doi = {10.21468/SciPostPhysCodeb.10},
	url = {https://scipost.org/10.21468/SciPostPhysCodeb.10}
}

@article{gray2018quimb,
  author  = {Gray, Johnnie},
  title   = {quimb: A {P}ython package for quantum information and many-body calculations},
  journal = {J. Open Source Softw.},
  volume  = {3},
  pages   = {819},
  year    = {2018},
}

@article{Mortier_2025,
   title={Fermionic tensor network methods},
   volume={18},
   ISSN={2542-4653},
   url={http://dx.doi.org/10.21468/SciPostPhys.18.1.012},
   DOI={10.21468/scipostphys.18.1.012},
   number={1},
   journal={SciPost Physics},
   publisher={Stichting SciPost},
   author={Mortier, Quinten and Devos, Lukas and Burgelman, Lander and Vanhecke, Bram and Bultinck, Nick and Verstraete, Frank and Haegeman, Jutho and Vanderstraeten, Laurens},
   year={2025},
   month=Jan }

@software{quantax,
  author  = {Chen, Ao and Roth, Christopher},
  title   = {Quantax: Flexible neural quantum states in JAX},
  year    = {2024},
  url     = {https://github.com/ChenAo-Phys/quantax},
  version = {0.2.1}
}

@software{Du_vmc_torch_Flexible_Variational,
author = {Du, Si-Jing},
license = {Apache-2.0},
year    = {2026},
title = {{vmc\_torch: Flexible Variational Monte Carlo for Quantum Many-Body Systems with PyTorch}},
url = {https://github.com/sjdu10/vmc_torch},
version = {0.1.0}
}

@misc{symmray,
  howpublished = {https://github.com/jcmgray/symmray},
  url     = {https://github.com/jcmgray/symmray}
}

@article{chen2024minsr,
  author  = {Chen, Ao and Heyl, Markus},
  title   = {Empowering deep neural quantum states through efficient optimization},
  journal = {Nat. Phys.},
  volume  = {20},
  pages   = {1476},
  year    = {2024},
  doi     = {10.1038/s41567-024-02566-1},
  eprint  = {2302.01941},
}

@article{sorella1998green,
   title={Green Function Monte Carlo with Stochastic Reconfiguration},
   volume={80},
   ISSN={1079-7114},
   url={http://dx.doi.org/10.1103/PhysRevLett.80.4558},
   DOI={10.1103/physrevlett.80.4558},
   number={20},
   journal={Physical Review Letters},
   publisher={American Physical Society (APS)},
   author={Sorella, Sandro},
   year={1998},
   month=May, pages={4558–4561} }

@article{liu2025fpeps,
  title = {Accurate Simulation of the Hubbard Model with Finite Fermionic Projected Entangled Pair States},
  author = {Liu, Wen-Yuan and Zhai, Huanchen and Peng, Ruojing and Gu, Zheng-Cheng and Chan, Garnet Kin-Lic},
  journal = {Phys. Rev. Lett.},
  volume = {134},
  issue = {25},
  pages = {256502},
  numpages = {8},
  year = {2025},
  month = {Jun},
  publisher = {American Physical Society},
  doi = {10.1103/r4q9-4yvj},
  url = {https://link.aps.org/doi/10.1103/r4q9-4yvj}
}

\end{document}